\documentclass[12pt, a4paper]{article}

\usepackage[T1]{fontenc}
\usepackage[utf8]{inputenc}
\usepackage{lmodern}

\usepackage[a4paper, top=22mm, bottom=25mm, left=25mm, right=25mm]{geometry}

\usepackage{graphicx}
\usepackage{amsmath,amssymb}
\usepackage{mathtools}
\usepackage{mathrsfs}
\usepackage{wasysym}
\usepackage{esint}
\usepackage{siunitx}
\usepackage{microtype}
\usepackage{booktabs}
\usepackage{array}
\usepackage{tabularx}
\usepackage{float}
\usepackage{caption}
\usepackage{subcaption}
\usepackage{xcolor}
\usepackage{hyperref}
\usepackage{url}
\usepackage{parskip}
\usepackage{titlesec}
\usepackage{enumitem}
\usepackage{natbib}
\allowdisplaybreaks

\hypersetup{
  colorlinks  = true,
  urlcolor    = blue,
  linkcolor   = blue,
  citecolor   = blue,
  pdftitle    = {Development and validation of a local neoclassical transport module in NLT with applications to EAST-relevant impurity transport and trapped-electron-mode stability},
  pdfauthor   = {Dandan Liao, Lei Ye, Guoxu Wang, Xiaotao Xiao, Chengkang Pan, Qilong Ren, Wei Zhang, Nong Xiang},
  pdfkeywords = {neoclassical transport, gyrokinetic simulation, multi-species collision, Sugama collision operator, impurity transport, trapped-electron mode}
}

\titleformat{\section}[block]
  {\normalfont\bfseries\normalsize}{\thesection.}{0.5em}{}
\titleformat{\subsection}[block]
  {\normalfont\bfseries\small}{\thesubsection.}{0.5em}{}
\titleformat{\subsubsection}[block]
  {\normalfont\itshape\small}{\thesubsubsection.}{0.5em}{}
\titlespacing*{\section}       {0pt}{9pt plus 2pt minus 1pt}{4pt}
\titlespacing*{\subsection}    {0pt}{7pt plus 2pt minus 1pt}{3pt}
\titlespacing*{\subsubsection} {0pt}{5pt plus 1pt minus 1pt}{2pt}

\begin{document}

{\fontsize{22}{27}\selectfont\bfseries
Development and validation of a local neoclassical transport module in NLT with applications to EAST-relevant impurity transport and trapped-electron-mode stability%
\par}

\vspace{14pt}

{\normalsize\bfseries
Dandan Liao\textsuperscript{1,2},
Lei Ye\textsuperscript{1,*},
Guoxu Wang\textsuperscript{1,2},
Xiaotao Xiao\textsuperscript{1},
Chengkang Pan\textsuperscript{1},
Qilong Ren\textsuperscript{1},
Wei Zhang\textsuperscript{1}
and Nong Xiang\textsuperscript{1}
}

\vspace{8pt}

{\small
\textsuperscript{1} Institute of Plasma Physics, Hefei Institutes of Physical Science, Chinese Academy of Sciences, Hefei, 230031, China\\[1pt]
\textsuperscript{2} University of Science and Technology of China, Hefei, 230031, China
}

\vspace{5pt}

{\small
\textsuperscript{*}E-mail of corresponding author(s):
\href{mailto:lye@ipp.ac.cn}{lye@ipp.ac.cn}
}

\vspace{6pt}


\vspace{8pt}
\noindent\rule{\linewidth}{0.6pt}
\vspace{4pt}

\noindent\textbf{Abstract}\\[2pt]
\noindent
A local neoclassical transport module has been developed and validated in the semi-Lagrangian gyrokinetic code NLT for multi-species collisional plasmas. The module incorporates a linearized multi-species Sugama collision operator and provides two complementary solution strategies. In the initial-value formulation, a composite substep source-integration scheme is introduced to accurately evaluate the neoclassical drive along unperturbed particle trajectories while retaining large macroscopic time steps. A direct steady-state solver is also implemented to obtain the stationary neoclassical response without long-time relaxation. The two approaches are benchmarked against the Eulerian neoclassical code NEO for electron-ion plasmas and three-species plasmas with carbon impurities. The NLT results reproduce the NEO particle and heat fluxes, parallel flows, and bootstrap current over a broad collisionality range. As representative applications, the validated framework is applied to EAST-relevant tungsten impurity transport and core trapped-electron-mode stability. The results show that tungsten neoclassical transport is sensitive to local profile gradients, while the increased effective collisionality associated with larger \(Z_{\rm eff}\) can reduce the linear TEM growth rate under the considered EAST-relevant conditions. These developments extend NLT toward realistic multi-species collisional transport simulations.

\vspace{6pt}


\vspace{5pt}

\vspace{6pt}
\noindent\rule{\linewidth}{0.6pt}
\vspace{10pt}

\section{Introduction}

Understanding cross-field transport in magnetically confined plasmas is essential for the prediction and optimization of fusion performance. In addition to turbulence-driven transport, neoclassical transport caused by the combined effects of Coulomb collisions and drift-orbit dynamics in toroidal geometry provides an important baseline for particle, momentum, and heat transport \cite{Hinton1985,Helander2002}. Although turbulent transport often dominates in the core plasma, neoclassical effects remain important in internal and edge transport barriers and are particularly relevant to heavy-impurity transport. Tungsten (W), a leading plasma-facing material for reactor-relevant devices such as ITER, can undergo strong neoclassical inward convection because of its large charge number, potentially leading to core accumulation~\cite{Hirshman1981}. Previous modeling studies have also shown that tungsten radiation loss is closely correlated with impurity transport levels~\cite{CFEDR_W2025}. Collisions may also modify turbulent transport by changing microinstability dynamics and turbulence-regulation mechanisms. For example, ion-ion collisions can damp zonal flows and thereby influence ion-temperature-gradient turbulence and heat transport \cite{Hinton1999}. These considerations motivate the development of gyrokinetic simulation tools with accurate and conservative multi-species collision models.

Local neoclassical transport has traditionally been described by drift-kinetic theory under the assumptions of axisymmetric magnetic geometry, prescribed equilibrium profiles, and scale separation between neoclassical and turbulent transport \cite{HH1972}. Based on this framework, the Eulerian drift-kinetic $\delta f$ code NEO, equipped with a fully linearized Fokker-Planck collision operator, has been widely used to provide accurate neoclassical transport calculations and benchmark references \cite{belli2008,belli2012,belli2015}. In parallel, gyrokinetic turbulence codes such as CGYRO have incorporated advanced collisional capabilities for local linear and nonlinear microturbulence simulations \cite{CandyCGYRO2016}. Global gyrokinetic codes, including XGC \cite{XGC2016,XGCneo2016,XGC2024}, ORB5 \cite{ORB52010,ORB52012,ORB52020}, GT5D \cite{GT5D2021}, and GYSELA \cite{GYSELA2018,GYSELA2023}, have further demonstrated that collisions, neoclassical dynamics, turbulence, radial electric fields, and impurity transport can interact in a non-additive manner. These studies show that reliable multi-species collision physics is not only necessary for neoclassical transport calculations, but also important for future integrated simulations of neoclassical and turbulent transport.

The semi-Lagrangian gyrokinetic code NLT has been developed for electromagnetic turbulence simulations in tokamak plasmas \cite{YE_2018}. In previous work, a multi-species linearized Sugama collision operator and a semi-implicit time-stepping scheme were implemented in the global version of NLT \cite{liao2026}. However, global neoclassical simulations are computationally expensive and, in the absence of profile sources, may not reach a well-defined quasi-steady state because the background profiles evolve under the resulting transport fluxes. For code verification, parameter scans, and experimental interpretation, a local neoclassical formulation with prescribed equilibrium parameters and thermodynamic gradients is therefore highly useful. This motivates the development of an efficient local neoclassical transport module in NLT.

In this work, we develop and validate a local neoclassical transport module in the semi-Lagrangian gyrokinetic code NLT for multi-species collisional plasmas. The module incorporates the linearized multi-species Sugama collision operator and supports two complementary solution strategies. The first is an initial-value relaxation approach, which is consistent with the time-evolution framework of NLT turbulence simulations. To improve the accuracy of large-time-step calculations, a composite substep source-integration scheme is introduced to evaluate the neoclassical drive along unperturbed particle trajectories. This scheme reduces source-integration errors associated with fast electron parallel motion while retaining the large macroscopic time steps enabled by the semi-implicit collision treatment. The second strategy is a direct steady-state solver, in which the stationary local drift-kinetic equation is solved as a linear system without following the long-time relaxation process. This approach is particularly useful for benchmark studies and collisionality scans.

The new local NLT module is benchmarked against the Eulerian neoclassical code NEO for both electron-ion plasmas and three-species plasmas containing carbon impurities. Particle and heat fluxes, parallel flows, and bootstrap current are compared over a broad range of collisionalities. After validation, the upgraded framework is applied to EAST-relevant conditions to study impurity neoclassical transport and the effects of Coulomb collisions and high-\(Z\) impurities on core trapped-electron-mode (TEM) stability. In the present work, the neoclassical transport calculations and the linear TEM simulations are performed independently. Therefore, the EAST-relevant results should be regarded as representative applications of the validated collisional NLT framework, while self-consistent coupling between impurity transport and turbulence is left for future work.

The remainder of this paper is organized as follows. Section~2 presents the local neoclassical model and the numerical algorithms, including the composite substep source-integration scheme and the direct steady-state solver. Section~3 reports benchmark comparisons with NEO for electron-ion and impurity-containing plasmas. Section~4 applies the validated framework to EAST-relevant impurity transport and TEM stability. Section~5 summarizes the main results and discusses future work.

\section{Local neoclassical transport model and numerical algorithms}

\subsection{Local neoclassical drift-kinetic equation}

In an axisymmetric tokamak equilibrium, the magnetic geometry is described by magnetic-flux coordinates \((\psi,\theta,\zeta)\), where \(\psi\) is the poloidal magnetic flux, \(\theta\) is the poloidal angle, and \(\zeta\) is the toroidal angle. Using straight-field-line magnetic coordinates, the equilibrium magnetic field can be written in the Clebsch form
\begin{align}
    \mathbf B     =     q(\psi)\nabla\psi\times\nabla\theta   +   \nabla\zeta\times\nabla\psi,
\end{align}
where \(q(\psi)\) is the safety factor. The NLT code uses the field-aligned coordinates \((x,y,z)\) \cite{YE_2018},
\begin{align}
    x=r(\psi), \qquad
    y=q(\psi)\theta-\zeta, \qquad
    z=\theta ,
\end{align}
where \(r(\psi)\) is a minor-radius-like flux coordinate. In these coordinates,
\begin{align}
    \mathbf B
    =
    \frac{d\psi}{dr}\nabla x\times\nabla y ,
\end{align}
and the corresponding spatial Jacobian is
\begin{align}
    J_x     =     \left(    \nabla x\times\nabla y\cdot\nabla z     \right)^{-1}.
\end{align}
Here, \(x\) labels the flux surface, \(y\) labels the field line on that surface, and \(z\) denotes the coordinate along the magnetic field line. In the local neoclassical model considered below, the calculation is performed on a prescribed flux surface, with fixed equilibrium parameters and fixed radial thermodynamic gradients.

    For species \(j\), the gyrocenter phase-space coordinate is denoted by \(\mathcal Z=(\mathbf X,v_\parallel,\mu)\), where \(\mathbf X=(x,y,z)\), \(v_\parallel\) is the parallel velocity, and \(\mu=m_j v_\perp^2/(2B)\) is the magnetic moment. The distribution function is decomposed as
\begin{align}
    f_j(\mathcal Z,t)     =     f_{Mj}(\mathcal Z)    +     \delta f_j(\mathcal Z,t),
\end{align}
where the local Maxwellian is
\begin{align}
    f_{Mj}     =    n_j    \left(    \frac{m_j}{2\pi T_j}    \right)^{3/2}    \exp\left[
    -\frac{m_jv_\parallel^2/2+\mu B}{T_j}     \right].
\end{align}
Here, \(n_j\) and \(T_j\) are flux-surface quantities. Their radial gradients provide the thermodynamic drive for neoclassical transport.

For an axisymmetric local neoclassical problem, the response is independent of the field-line label \(y\). Consistent with the local drift-kinetic ordering, the lowest-order characteristics are restricted to a fixed flux surface, while the radial magnetic drift enters as a source term through the radial gradients of the Maxwellian. The evolution equation for \(\delta f_j\) can then be written as \cite{belli2008}
\begin{align}
    \left.    \frac{d\delta f_j}{dt}     \right|_{\rm NC}     =     \mathcal S_{j,{\rm neo}}  +     C_j(\delta f),
    \label{eq:local_df}
\end{align}
with
\begin{align}
    \left.     \frac{d}{dt}     \right|_{\rm NC}     &\equiv     \frac{\partial}{\partial t}     +     \mathcal L_{\rm NC},
    \\
    \mathcal L_{\rm NC}     &\equiv     \dot z     \frac{\partial}{\partial z}     +     \dot v_\parallel     \frac{\partial}{\partial v_\parallel}.
    \label{eq:L_nc}
\end{align}
The local characteristics are
\begin{align}
    \dot z     &=    \frac{d\psi}{dr}     \frac{v_\parallel}{J_x B_{\parallel j}^*}, 
    \label{eq:zdot_local}
    \\
    \dot v_\parallel     &=    -     \frac{d\psi}{dr}     \frac{\mu}{m_j J_x B_{\parallel j}^*}     \frac{dB}{dz},
    \label{eq:vdot_local}
\end{align}
where
\begin{align}
    \mathbf B_j^*     =    \mathbf B    +     \frac{m_j v_\parallel}{Z_j e}     \nabla\times\hat{\mathbf b}, 
    \qquad
    B_{\parallel j}^*     =    \hat{\mathbf b}\cdot\mathbf B_j^* . 
    \label{eq:Bstar}
\end{align}
Here, \(\hat{\mathbf b}=\mathbf B/B\), and \(Z_j\) is the charge number of species \(j\). 

The magnetic drift velocity entering the neoclassical source term is
\begin{align}
    \mathbf v_{Dj}     =     \frac{1}{\Omega_j}     \hat{\mathbf b}\times     \left(
    v_\parallel^2\boldsymbol{\kappa}     +     \frac{\mu}{m_j}\nabla B     \right),
    \label{eq:vD}
\end{align}
where \(\Omega_j=Z_j eB/m_j\) is the gyrofrequency and \(\boldsymbol{\kappa}=\hat{\mathbf b}\cdot\nabla\hat{\mathbf b}\) is the magnetic-field curvature. The neoclassical source term is driven by the radial magnetic drift across the equilibrium density and temperature gradients:
\begin{align}
    \mathcal S_{j,{\rm neo}}     =     -    \left(    \mathbf v_{Dj}\cdot\nabla x     \right)     \left[     \frac{\partial_x n_j}{n_j}     +     \left(    \frac{m_jv_\parallel^2/2+\mu B}{T_j}     -     \frac{3}{2}    \right)     \frac{\partial_x T_j}{T_j}     \right]     f_{Mj}.
    \label{eq:sneo}
\end{align}
Here, \(\partial_x\) denotes the radial derivative with respect to the flux-surface label \(x\).

The collision term is modeled using the linearized multi-species Sugama collision operator. In compact form,
\begin{align}
    C_j(\delta f)     =     \sum_{j'}     C^{\rm S}_{jj'}(\delta f_j,\delta f_{j'}), 
    \label{eq:sugama_compact}
\end{align}
where the summation includes both self-collisions and inter-species collisions. The operator contains the test-particle part, the field-particle restoring terms, and the non-isothermal correction terms. It conserves particle number for each species and total momentum and energy for each colliding species pair, while preserving self-adjointness. These properties are important for obtaining physically consistent neoclassical fluxes, parallel flows, and bootstrap current in multi-species plasmas. The detailed form of the Sugama operator and its implementation in NLT have been described in Refs.~\cite{sugama2009,liao2026} and are not repeated here.

\subsection{Initial-value algorithm with composite substep source integration}

The initial-value approach is used to maintain consistency with the time-evolution framework of NLT turbulence simulations. This consistency is useful for future self-consistent simulations in which neoclassical and turbulent transport are evolved within the same gyrokinetic framework \cite{ORB52012}. In the semi-Lagrangian formulation, Eq.~\eqref{eq:local_df} is advanced by integrating along the local collisionless characteristics. The evolution contains two parts: the source-driven characteristic update and the collisional relaxation. These two parts are separated using a Strang splitting scheme \cite{strang1968}. Over one macroscopic time step \(\Delta t\), the update is written as
\begin{align}
    \delta f^{n+1}    =     \mathcal U_S\left(\frac{\Delta t}{2}\right)     \mathcal U_C\left(\Delta t\right)     \mathcal U_S\left(\frac{\Delta t}{2}\right)     \delta f^n.
    \label{eq:strang_split}
\end{align}
Here, \(\mathcal U_C(\tau)\) denotes the collisional relaxation operator, \(\partial\delta f_j/\partial t=C_j(\delta f)\), which is advanced using the semi-implicit multi-species collision scheme developed in Ref.~\cite{liao2026}. The collisional update is therefore not repeated here; instead, we focus on the source-update operator \(\mathcal U_S\). The operator \(\mathcal U_S(\tau)\) corresponds to the collisionless source problem governed by the local neoclassical characteristic derivative,
\begin{align}
    \left.\frac{d\delta f}{dt}\right|_{\rm NC}
    =
    \mathcal S_{\rm neo}.
    \label{eq:source_characteristic}
\end{align}
Here, \(\left.d/dt\right|_{\rm NC}\) denotes the derivative along the local neoclassical characteristics defined by Eqs.~\eqref{eq:zdot_local} and \eqref{eq:vdot_local}. For a source half step from \(t^n\) to \(t^{n+1/2}\), consider a fixed Eulerian phase-space grid point \(\mathcal Z_I\) at \(t^{n+1/2}\). Tracing the characteristic backward to its foot point \(\mathcal Z_I^*\) at \(t^n\), the formal semi-Lagrangian solution is
\begin{align}
    \delta f(\mathcal Z_I,t^{n+1/2})
    =
    \delta f(\mathcal Z_I^*,t^n)
    +
    \int_{t^n}^{t^{n+1/2}}
    \mathcal S_{\rm neo}\bigl(\mathcal Z_{\rm NC}(t)\bigr)\,dt .
    \label{eq:source_formal_solution}
\end{align}
If the same macroscopic time step is used for both the trajectory integration and the source quadrature, the source integral may be inaccurate when \(\Delta t\) is much larger than the electron bounce or transit time. In that case, the time-step restriction is not caused by the stability of the semi-implicit collision update, but by the accuracy of the orbit-integrated neoclassical source term, especially for thermal electrons.  For the local problem considered here, the equilibrium profiles and their radial gradients are prescribed and time independent. Therefore, \(\mathcal S_{\rm neo}\) has no explicit time dependence, and the source increment over a given macroscopic time step can be precomputed. To improve the accuracy of this orbit-integrated source while retaining a large macroscopic time step, we introduce a composite substep source-integration scheme.

Let \(N_{\rm orb}\) be the number of orbit substeps associated with one macroscopic time step \(\Delta t\). For compactness, the species index is omitted in this description, although different species-dependent values of \(N_{\rm orb}\) can be used in practice. The orbit substep is
\begin{align}
    \Delta\tau     =    \frac{\Delta t}{N_{\rm orb}} .
\end{align}
Since the source update in Eq.~\eqref{eq:strang_split} is performed over a half step, each source half step contains
\begin{align}
    N_{1/2}     =    \frac{N_{\rm orb}}{2}
\end{align}
substeps, where \(N_{\rm orb}\) is chosen to be even.

For the first source half step, the backward time nodes are defined as
\begin{align}
    t_k     =     t^{n+1/2}     -     k\Delta\tau,     \qquad     k=0,1,\cdots,N_{1/2},
\end{align}
so that \(t_0=t^{n+1/2}\) and \(t_{N_{1/2}}=t^n\). The corresponding phase-space locations are denoted by \(\mathcal Z_k=\mathcal Z(t_k)\), with \(\mathcal Z_0=\mathcal Z_I\) and \(\mathcal Z_{N_{1/2}}=\mathcal Z_I^*\). The source integral is decomposed as
\begin{align}
    \int_{t^n}^{t^{n+1/2}}     \mathcal S_{\rm neo}\bigl(\mathcal Z(t)\bigr)\,dt     =     \sum_{k=0}^{N_{1/2}-1}     \int_{t_{k+1}}^{t_k}    \mathcal S_{\rm neo}\bigl(\mathcal Z(t)\bigr)\,dt . 
\end{align}
Using a composite trapezoidal rule along the substepped orbit gives
\begin{align}
    \int_{t^n}^{t^{n+1/2}} \mathcal S_{\rm neo}\bigl(\mathcal Z(t)\bigr)\,dt
    &\simeq
    \Delta\tau \biggl[
    \frac{\mathcal S_{\rm neo}(\mathcal Z_0)
    + \mathcal S_{\rm neo}(\mathcal Z_{N_{1/2}})}{2}
    \nonumber\\
    &\quad + \sum_{k=1}^{N_{1/2}-1}
    \mathcal S_{\rm neo}(\mathcal Z_k) \biggr].
    \label{eq:composite_source}
\end{align}
The second source half step, from \(t^{n+1/2}\) to \(t^{n+1}\), is evaluated after the collisional update using the same composite source integration.
    In the NLT implementation, the foot points and source increments associated with the source half step are computed once during initialization for each Eulerian phase-space grid point. The pre-integrated source increment is then stored on the phase-space grid and reused during the subsequent time advancement of \(\delta f\). Therefore, the additional cost of the composite source integration is mainly introduced during initialization rather than at every macroscopic time step. If the macroscopic time step, the number of orbit substeps, or the equilibrium profiles are changed, the source increments are recomputed.

    The value of \(N_{\rm orb}\) is chosen according to the macroscopic time step and the relevant orbital time scales. A larger \(N_{\rm orb}\) is required when \(\Delta t\) is large compared with the electron bounce or transit time, or in low-collisionality cases where the initial-value simulation must follow the neoclassical relaxation over many orbital periods. By resolving the variation of the neoclassical source along unperturbed characteristics, the composite substep scheme reduces trajectory and quadrature errors without reducing the macroscopic time step. This allows the large-step advantage of the semi-implicit collision scheme to be retained more effectively in low-collisionality neoclassical simulations.

\subsection{Direct steady-state solver}

As a complementary approach to the initial-value method, the local neoclassical equation can also be solved directly in its steady-state form. Setting only the explicit time derivative in Eq.~\eqref{eq:local_df} to zero, while retaining the phase-space advection operator \(\mathcal L_{\rm NC}\), gives
\begin{align}
    \mathcal L_{\rm NC}\delta f_j = \mathcal S_{j,{\rm neo}} + C_j(\delta f).
    \label{eq:steady_df}
\end{align}
This formulation computes the stationary neoclassical response directly, rather than following the physical relaxation process in time. It is therefore useful for benchmark calculations and parameter scans.

Using the characteristic coefficients defined above, Eq.~\eqref{eq:steady_df} can be written explicitly as
\begin{align}
    \frac{d\psi}{dr}\frac{v_\parallel}{J_x B_{\parallel j}^*}\frac{\partial \delta f_j}{\partial z}
    -\frac{d\psi}{dr}\frac{\mu}{m_j J_x B_{\parallel j}^*}\frac{dB}{dz}\frac{\partial \delta f_j}{\partial v_\parallel}
    - C_j(\delta f)
    =
    \mathcal S_{j,{\rm neo}} .
    \label{eq:direct_steady_state}
\end{align}
This equation represents the stationary balance among parallel streaming, mirror-force acceleration, collisional relaxation, and the neoclassical thermodynamic drive.

After discretization in \((z,v_\parallel,\mu)\) for all species, Eq.~\eqref{eq:direct_steady_state} is transformed into a coupled linear algebraic system,
\begin{align}
    \mathbf A\mathbf x = \mathbf b .
    \label{eq:linear_system}
\end{align}
The unknown vector \(\mathbf x\) contains the discrete values of \(\delta f_j\) at all phase-space grid points for all species. If \(n_z\), \(n_{v_\parallel}\), \(n_\mu\), and \(n_s\) denote the numbers of grid points in \(z\), \(v_\parallel\), \(\mu\), and species, respectively, the total number of unknowns is
\begin{align}
    n_g = n_z n_{v_\parallel} n_\mu n_s .
\end{align}
The right-hand side \(\mathbf b\) is assembled from the neoclassical source term \(\mathcal S_{j,{\rm neo}}\).

The matrix operator is decomposed as
\begin{align}
    \mathbf A = \mathbf A_z + \mathbf A_{v_\parallel} - \mathbf C ,
    \label{eq:A_decomp}
\end{align}
where \(\mathbf A_z\) represents the parallel-streaming term, \(\mathbf A_{v_\parallel}\) represents the mirror-force term, and \(\mathbf C\) is the discretized multi-species Sugama collision operator. In the present implementation, the advection terms in \(z\) and \(v_\parallel\) are approximated using fourth-order central finite differences, while the Sugama collision operator is discretized using the second-order finite-volume scheme developed in Ref.~\cite{liao2026}. Periodicity is imposed in the poloidal coordinate \(z\). The velocity-space domain is chosen sufficiently large so that the contribution from the Maxwellian tail at the boundary is negligible; the velocity-space truncation used in the benchmark calculations is specified in Sec.~3.

The collision operator contains both local velocity-space diffusion and nonlocal moment couplings. In particular, the field-particle restoring terms couple different velocity-space grid points through density, momentum, and energy moments, while inter-species collisions couple different species through momentum and energy exchange. Therefore, the resulting matrix contains sparse advection-like components together with moment-coupled collision components. Direct factorization of the full matrix is computationally expensive for practical phase-space resolutions. In the present implementation, the linear system is solved through the PETSc library. For the benchmark-size systems considered in this work, a parallel sparse direct solver based on MUMPS is used to obtain a robust reference solution. The code also supports a restarted-GMRES solver with right preconditioning for larger systems, in which a block-Jacobi preconditioner is combined with local LU solves on each subdomain. When it is used, the relative residual tolerance is typically \(10^{-12}\), and the true residual is monitored to verify the linear-solver accuracy.

In the local \(\delta f\) formulation used here, the background Maxwellian \(f_{Mj}\) is prescribed by the fixed flux-surface density, temperature, and their radial gradients. The direct solver computes the corresponding stationary nonadiabatic neoclassical response \(\delta f_j\) directly from the discretized steady drift-kinetic equation. The accuracy of the computed response is assessed through benchmark comparisons with NEO in Sec.~3.

The direct steady-state solver avoids long-time relaxation and repeated characteristic tracing. It is therefore computationally efficient for obtaining stationary neoclassical particle fluxes, heat fluxes, parallel flows, and bootstrap current over broad parameter ranges.

\section{Benchmark tests of the local neoclassical transport module}

This section presents benchmark tests of the newly developed local neoclassical transport module in NLT. The Eulerian drift-kinetic code NEO \cite{belli2008,belli2012,belli2015} is used as the reference code. The validation is designed to assess both the local neoclassical formulation and the numerical robustness of the two solution strategies introduced above. For electron-ion plasmas, both the initial-value relaxation approach and the direct steady-state solver are tested, providing two independent routes to the same steady neoclassical response. The initial-value calculation verifies the time-relaxation framework, including the semi-implicit collision treatment and the composite substep source-integration scheme, while the direct solver tests the discretized steady drift-kinetic formulation without relying on long-time relaxation.

The benchmark is further extended to a three-species impurity case with electrons, deuterium ions, and fully stripped carbon impurities. Using the direct steady-state solver, we compare the calculated particle fluxes, heat fluxes, parallel flows, and bootstrap current with NEO results over a broad collisionality range, providing a stringent test of the multi-species collision model.

\subsection{Benchmark setup and diagnostic quantities}

Unless otherwise specified, the benchmark calculations use the General Atomics standard case parameters \cite{GA1997}, as listed in Table~\ref{tab:ga_parameters}. This case has been widely used for local neoclassical code verification. The normalized ion gyroradius is set to \(\rho^*=1/1000\), so that the calculation remains within the local neoclassical ordering.

\begin{table}[htbp]
    \centering
    \caption{General Atomics standard case parameters \cite{GA1997}.}
    \label{tab:ga_parameters}
    \renewcommand{\arraystretch}{1.2}
    \begin{tabular}{lclc}
        \toprule
\(r/a=0.5\) & \(R/a=3.0\) \\
        \(q=2.0\) & \(\hat{s}=1.0\) \\
        \(a/L_n=1.0\) & \(T_i/T_e=1.0\) \\
        \(a/L_{Te}=a/L_{Ti}=3.0\) & \(m_i/m_e=3672\) \\\bottomrule
    \end{tabular}
\end{table}

The neoclassical particle flux, heat flux, \(B\)-weighted parallel flow, and \(B\)-weighted bootstrap current are evaluated as
\begin{align}
    \Gamma_j^{\rm neo} &= \left\langle \int d^3v\, v_{Dj}^r \delta f_j \right\rangle, \label{eq:Gamma_neo}\\
    Q_j^{\rm neo} &= \left\langle \int d^3v\, \frac{1}{2}m_jv^2 v_{Dj}^r \delta f_j \right\rangle, \label{eq:Q_neo}\\
    \langle U_{\parallel j}B\rangle &= \frac{1}{n_{0j}}\left\langle B\int d^3v\, v_\parallel \delta f_j \right\rangle, \label{eq:UparB}\\
    \langle j_\parallel B\rangle &= \sum_j Z_j e n_{0j}\langle U_{\parallel j}B\rangle . \label{eq:JparB}
\end{align}
Here, \(v_{Dj}^r=\mathbf v_{Dj}\cdot\nabla r\), \(v^2=v_\parallel^2+v_\perp^2\), and \(\langle\cdots\rangle\) denotes the flux-surface average. The \(B\)-weighted moments \(\langle U_{\parallel j}B\rangle\) and \(\langle j_\parallel B\rangle\) are used because they are the conventional flux-surface-averaged quantities reported in neoclassical theory. The corresponding flux-surface-averaged bootstrap current density can be obtained as \(j_{\rm bs}=\langle j_\parallel B\rangle/\langle B^2\rangle\).

All collisionality scans for the benchmark comparisons with the NEO code in the present work are performed using the normalized ion collision frequency defined in Ref.~\cite{belli2008}.
\begin{align}
    \frac{a}{v_{ti}}\tau_{ii}^{-1} = \frac{\sqrt{2}\pi n_i q_i^4\ln\Lambda\, a}{T_i^2},
    \label{eq:normalized_nuii}
\end{align}
where \(n_i\), \(q_i\), and \(T_i\) are the ion density, charge, and temperature, respectively, \(v_{ti}=\sqrt{T_i/m_i}\), and \(\ln\Lambda\) is the Coulomb logarithm.

\subsection{Electron-ion benchmark}

We first consider an electron-ion plasma using the parameters in Table~\ref{tab:ga_parameters}. Previous GENE-NEO neoclassical benchmarks mainly used a direct steady-state solver to obtain the stationary neoclassical response \cite{Doerk2012_big,CrandallGENE2020}. In the present work, this benchmark is used to test both the initial-value relaxation method and the direct steady-state solver, providing two complementary validations of electron-ion neoclassical transport in the local NLT module.

For the initial-value simulations, the phase-space resolution is \(N_z\times N_{v_\parallel}\times N_\mu=32\times32\times128\). The velocity-space domain is chosen as \(|v_\parallel|\leq 4.2v_T\) and \(v_\perp\leq 4.2v_T\), where \(v_T=\sqrt{T_0/m}\). The relatively high \(\mu\)-space resolution is used to reduce errors near the trapped-passing boundary and to maintain sufficient accuracy in the velocity-space moments.

\begin{figure}[t]
\centering
    \includegraphics[width=0.4\textwidth]{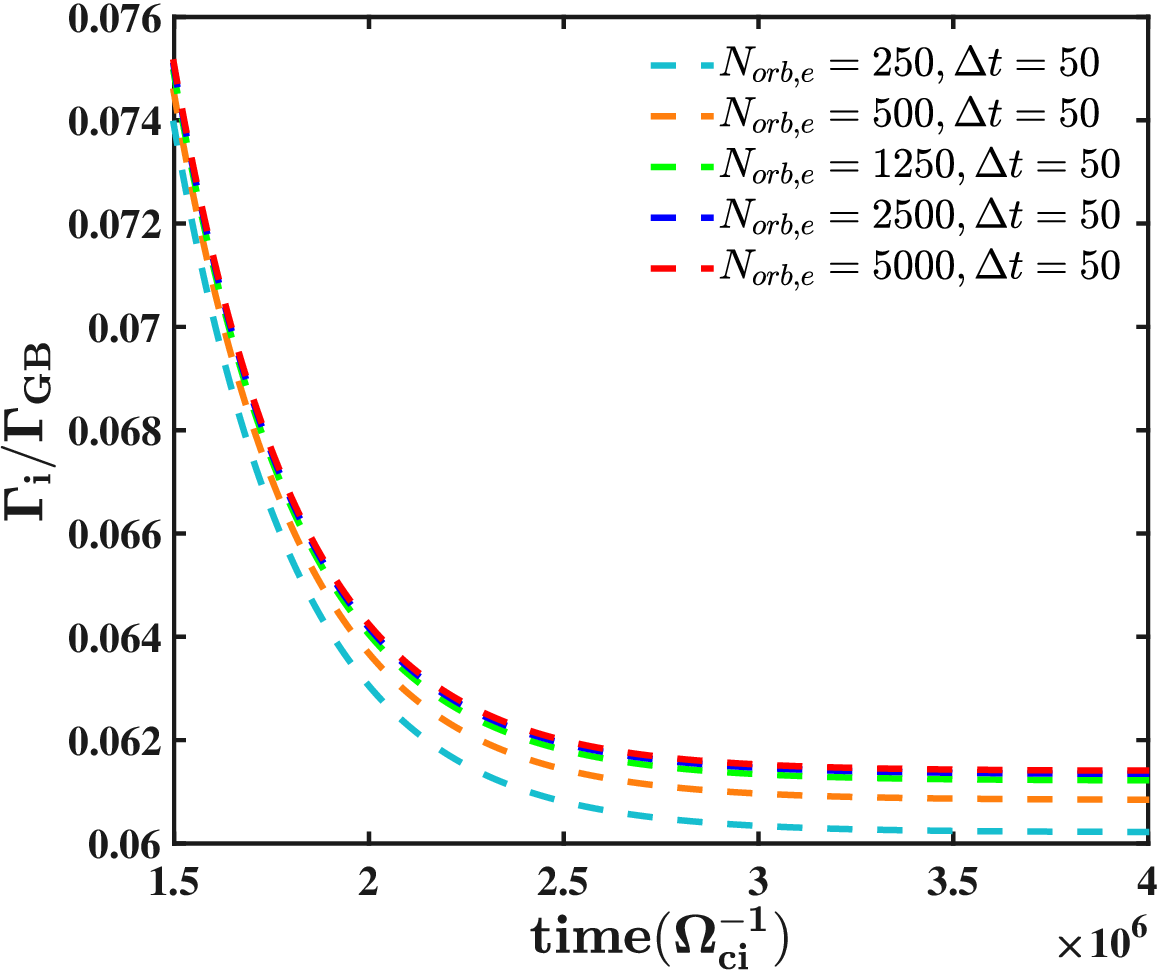}
    \includegraphics[width=0.4\textwidth]{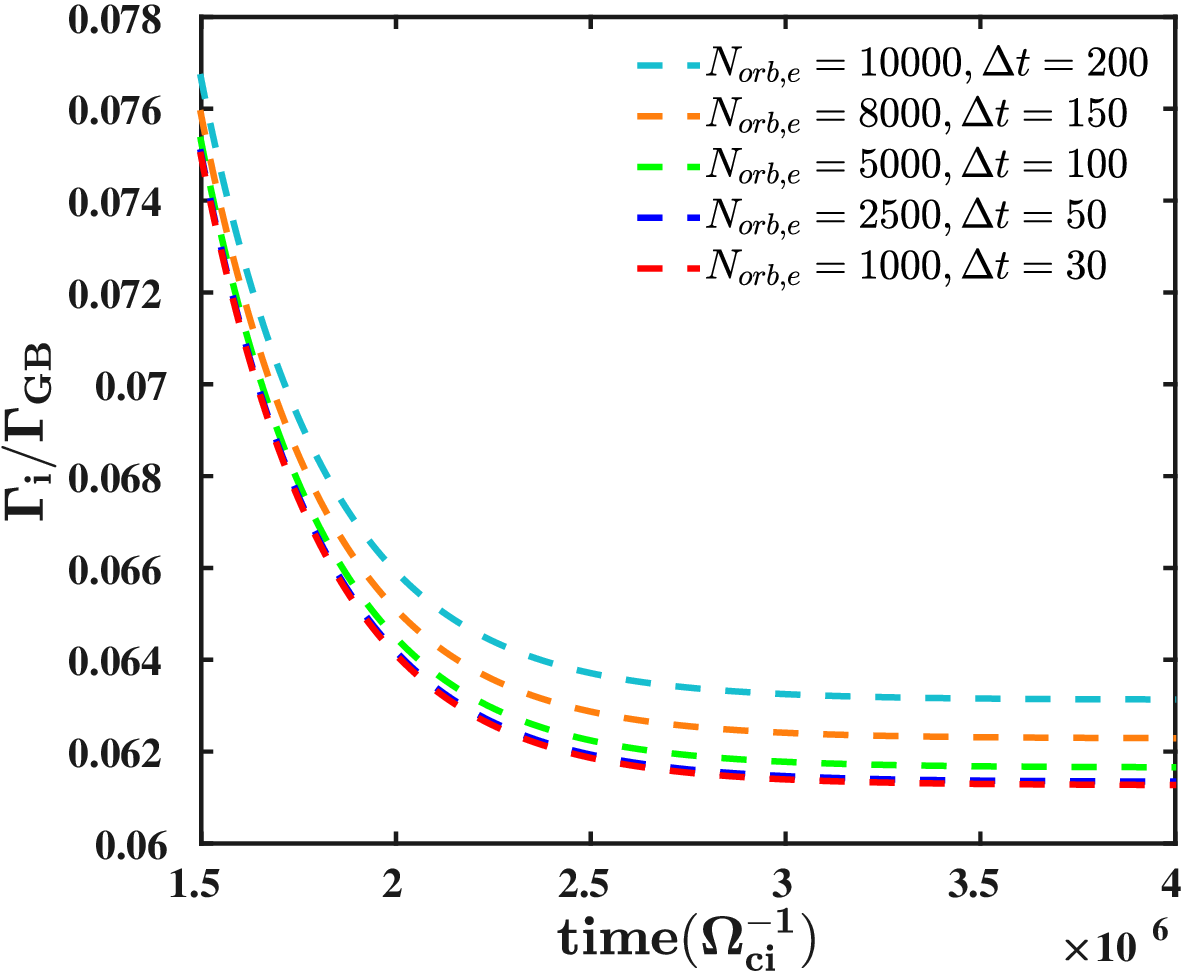}
    \caption{Numerical convergence of the composite substep source-integration scheme at \((a/v_{ti})\tau_{ii}^{-1}=3\times10^{-2}\). The left panel shows the convergence with \(N_{{\rm orb},e}\) at fixed \(\Delta t\), while the right panel shows the dependence on \(\Delta t\) with nearly fixed electron orbit substep size.}
    \label{fig:mstep_dt}
\end{figure}

We first examine the convergence of the composite substep source-integration scheme at \((a/v_{ti})\tau_{ii}^{-1}=3\times10^{-2}\). Since electrons have much faster parallel motion than ions, a larger number of orbit substeps is used for electrons. Unless otherwise specified, \(N_{{\rm orb},e}=5N_{{\rm orb},i}\), and \(N_{{\rm orb},e}\) is used to label the substep scan. Figure~\ref{fig:mstep_dt} shows the time evolution of the ion particle flux under two complementary scans. In the left panel, the macroscopic time step is fixed at \(\Delta t=50\Omega_{ci}^{-1}\), while \(N_{{\rm orb},e}\) is varied. The relaxation curves and the final steady particle flux converge as \(N_{{\rm orb},e}\) increases, indicating that the source-integration error is reduced by resolving the orbit more accurately. In the right panel, \(\Delta t\) is varied while \(N_{{\rm orb},e}\) is increased approximately in proportion to \(\Delta t\), so that the orbit substep size remains nearly fixed. The steady flux remains nearly unchanged for moderately large time steps, showing that the composite source integration allows a large macroscopic time step to be used without losing the accuracy of the source integral. A visible deviation appears when \(\Delta t\) becomes too large, indicating that the global splitting and time-discretization errors still impose a practical upper limit on the macroscopic time step.
\begin{figure}[t]
\centering
    \includegraphics[width=0.4\textwidth]{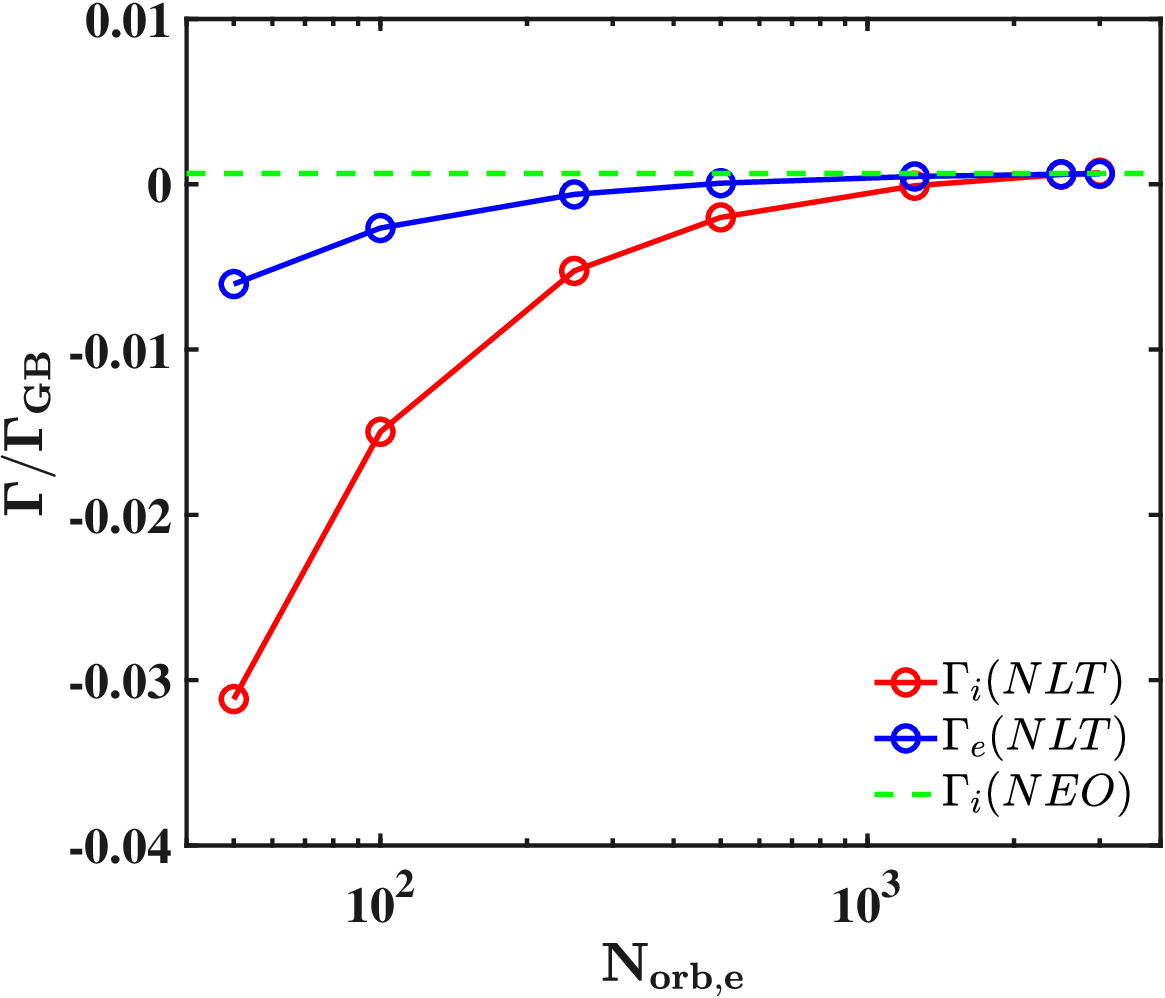}
    \caption{Dependence of the electron and ion particle fluxes on the substep number \(N_{\rm orb,e}\) in the composite substep source-integration scheme at \((a/v_{ti})\tau_{ii}^{-1}=3\times10^{-4}\) and \(\Delta t_{\max}=250\Omega_{ci}^{-1}\). The dashed horizontal line denotes the NEO reference value. Increasing \(N_{\rm orb,e}\) reduces the mismatch between \(\Gamma_i\) and \(\Gamma_e\), indicating improved recovery of intrinsic ambipolarity.}
    \label{fig:mstep_amb}
\end{figure}

A more stringent test is performed in the low-collisionality regime, \((a/v_{ti})\tau_{ii}^{-1}=3\times10^{-4}\), where the physical relaxation time is long and the initial-value calculation is more sensitive to accumulated orbit-integration errors. In an axisymmetric local neoclassical calculation, intrinsic ambipolarity provides a sensitive macroscopic diagnostic of numerical accuracy. Figure~\ref{fig:mstep_amb} shows the dependence of the electron and ion particle fluxes on the orbit substep number at \(\Delta t_{\max}=250\Omega_{ci}^{-1}\). When the orbit integration is insufficiently resolved, the ion and electron particle fluxes show a noticeable mismatch. As \(N_{{\rm orb},e}\) increases, both fluxes converge toward the NEO reference level and approach each other, demonstrating improved recovery of intrinsic ambipolarity.

\begin{figure}[htbp]
\centering
    \includegraphics[width=0.4\textwidth]{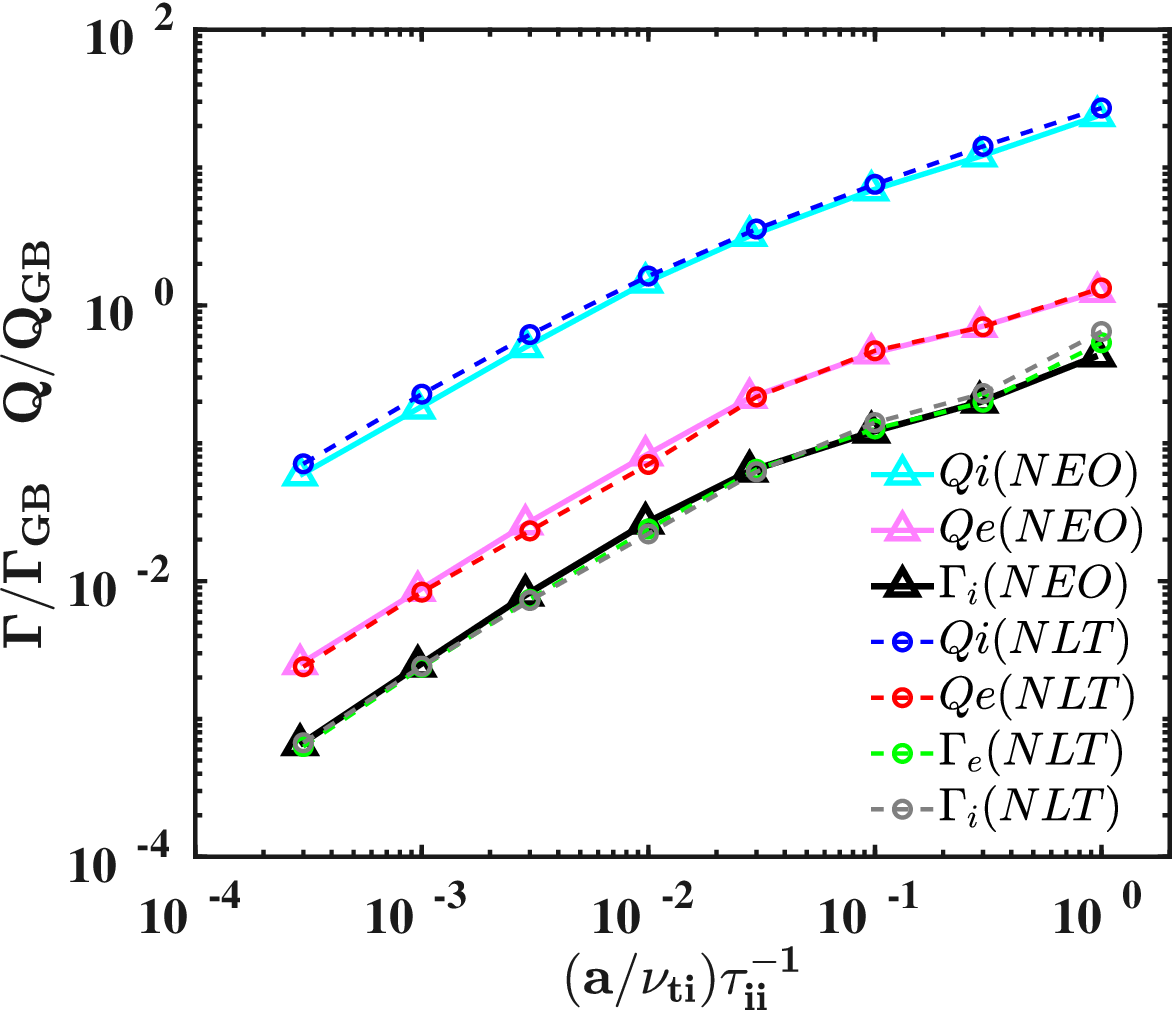}
    \includegraphics[width=0.43\textwidth]{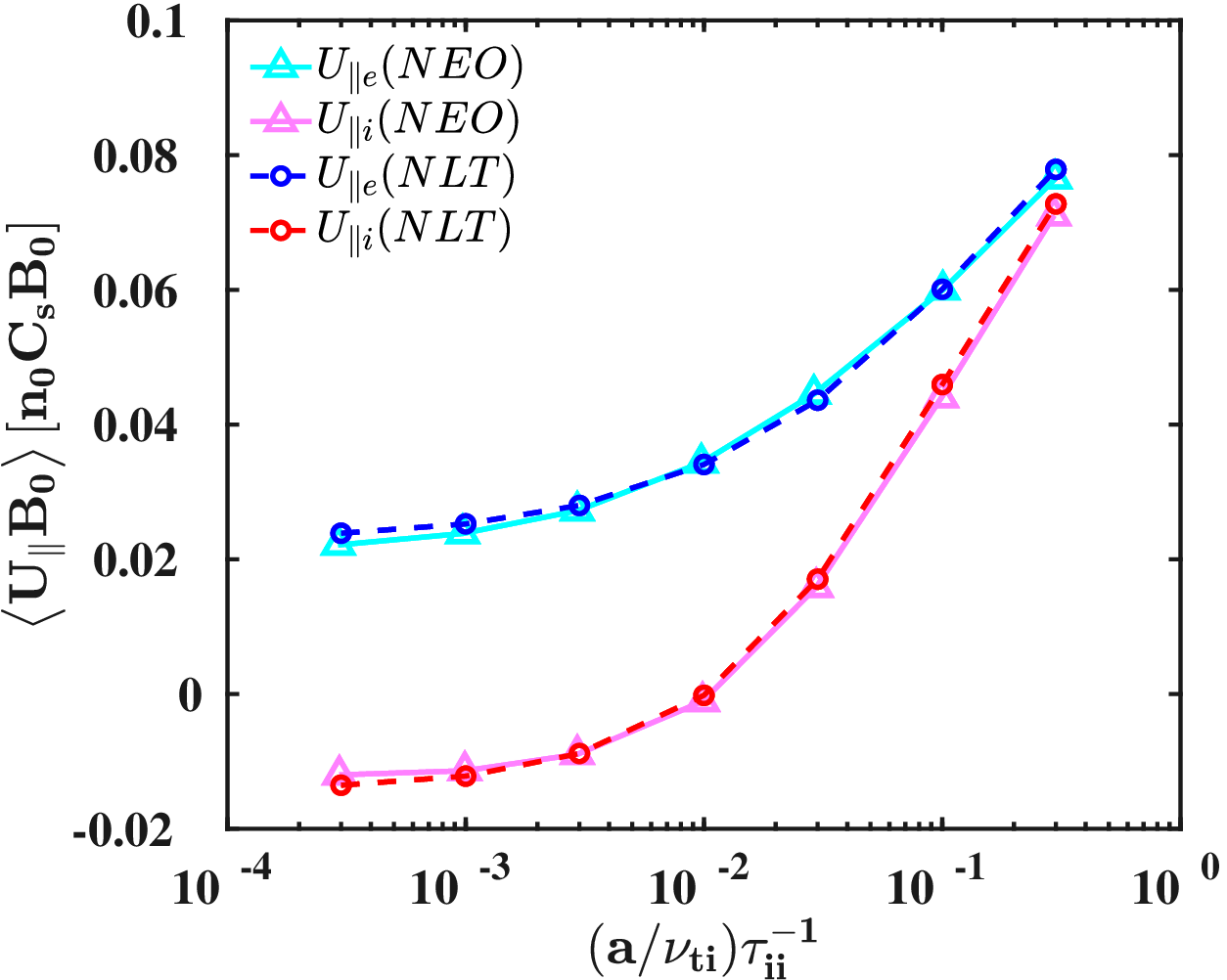}
    \caption{Benchmark of the initial-value relaxation solver against NEO. The left panel shows the electron and ion particle and heat fluxes, and the right panel shows the corresponding flux-surface-averaged parallel flows. Solid lines denote NEO results, and dashed lines denote NLT results.}
    \label{fig:neo_ei_iv}
\end{figure}

After establishing the convergence of the source integration, we compare the initial-value NLT results with NEO over a broad collisionality range. Figure~\ref{fig:neo_ei_iv} shows the particle and heat fluxes, together with the flux-surface-averaged parallel flows. The NLT results reproduce both the magnitude and the collisionality dependence of the NEO results for both species. The particle fluxes also satisfy intrinsic ambipolarity within the numerical accuracy of the calculation. This confirms that the initial-value formulation can recover the steady local neoclassical response through time relaxation. Over most of the scanned collisionality range, the discrepancies between NLT and NEO remain small for the plotted transport moments, except near sign-changing points where relative errors become ill-defined.

\begin{figure}[htbp]
\centering
    \includegraphics[width=0.4\textwidth]{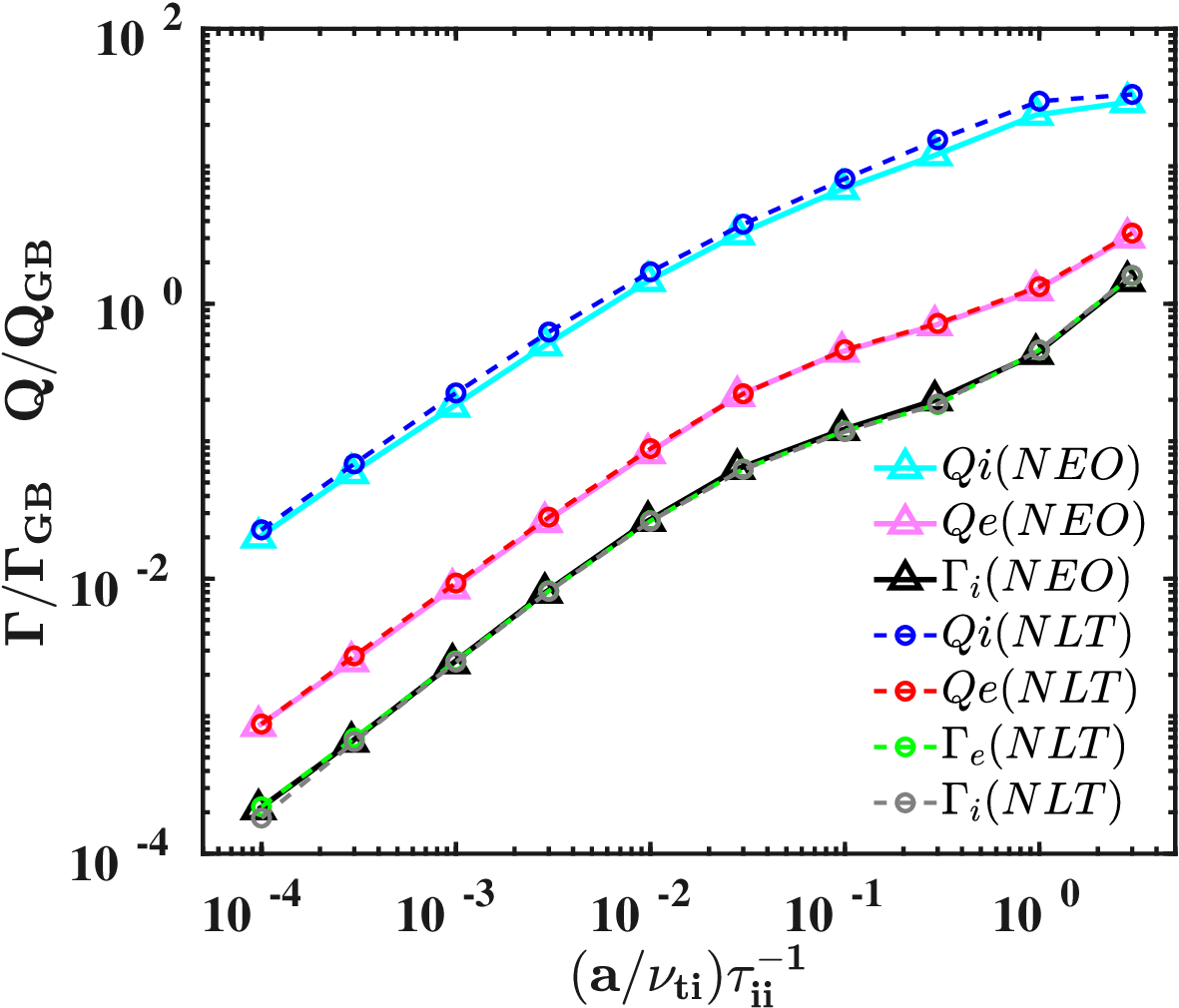}
    \includegraphics[width=0.43\textwidth]{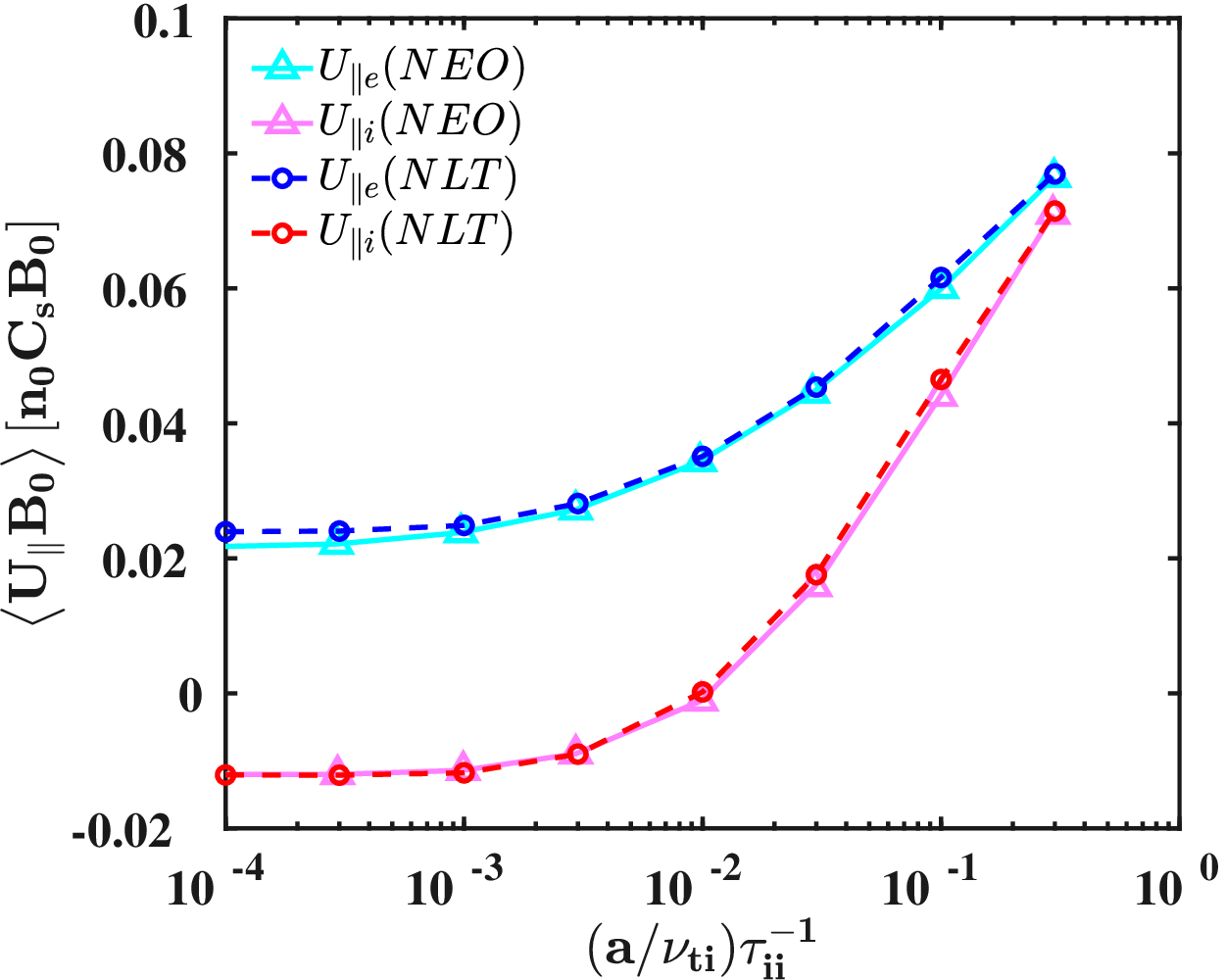}
    \caption{Benchmark of the direct steady-state solver against NEO. The left panel shows the electron and ion particle and heat fluxes, and the right panel shows the corresponding flux-surface-averaged parallel flows. Solid lines denote NEO results, and dashed lines denote NLT results.}
    \label{fig:neo_ei}
\end{figure}

The same benchmark is then repeated using the direct steady-state solver. Unlike the initial-value method, the direct solver obtains the stationary response from the discretized steady drift-kinetic equation without temporal relaxation or repeated characteristic tracing. The phase-space resolution is \(N_z\times N_{v_\parallel}\times N_\mu=32\times32\times32\). The corresponding results are shown in Fig.~\ref{fig:neo_ei}. The particle fluxes, heat fluxes, and parallel flows from the direct solver agree well with the NEO reference over the scanned collisionality range, demonstrating that the stationary formulation gives the same neoclassical response as the time-relaxation approach.

The different choices of \(N_\mu\) in the two approaches reflect their different numerical error mechanisms. In the initial-value formulation, the solution is obtained after long-time relaxation with repeated characteristic tracing, making the result more sensitive to the trapped-passing boundary and to accumulated orbit-integration errors. Therefore, a higher \(\mu\)-space resolution is used. In contrast, the direct steady-state solver computes the asymptotic response without repeated orbit tracing. For the macroscopic moments considered here, \(N_\mu=32\) is sufficient to reproduce the NEO fluxes and parallel flows. Taken together, the initial-value and direct-solver benchmarks provide two independent validations of the local NLT module for electron-ion neoclassical transport.

\subsection{Three-species benchmark with carbon impurities}

To further test the multi-species capability of the local NLT module, we consider a three-species plasma consisting of electrons, deuterium ions, and fully stripped carbon impurities. This case is solved using the direct steady-state solver. Compared with the electron-ion benchmark, this test is more stringent because it includes multiple self- and inter-species collision channels and requires consistent momentum and energy exchange among different species. The same GA standard parameters in Table~\ref{tab:ga_parameters} are used. The carbon impurity charge fraction is set to \(f_I=Z_I n_{0I}/n_{0e}=0.1\), with \(Z_I=6\), and the main-ion density is determined from quasineutrality.

\begin{figure}[htbp]
\centering
    \includegraphics[width=0.4\textwidth]{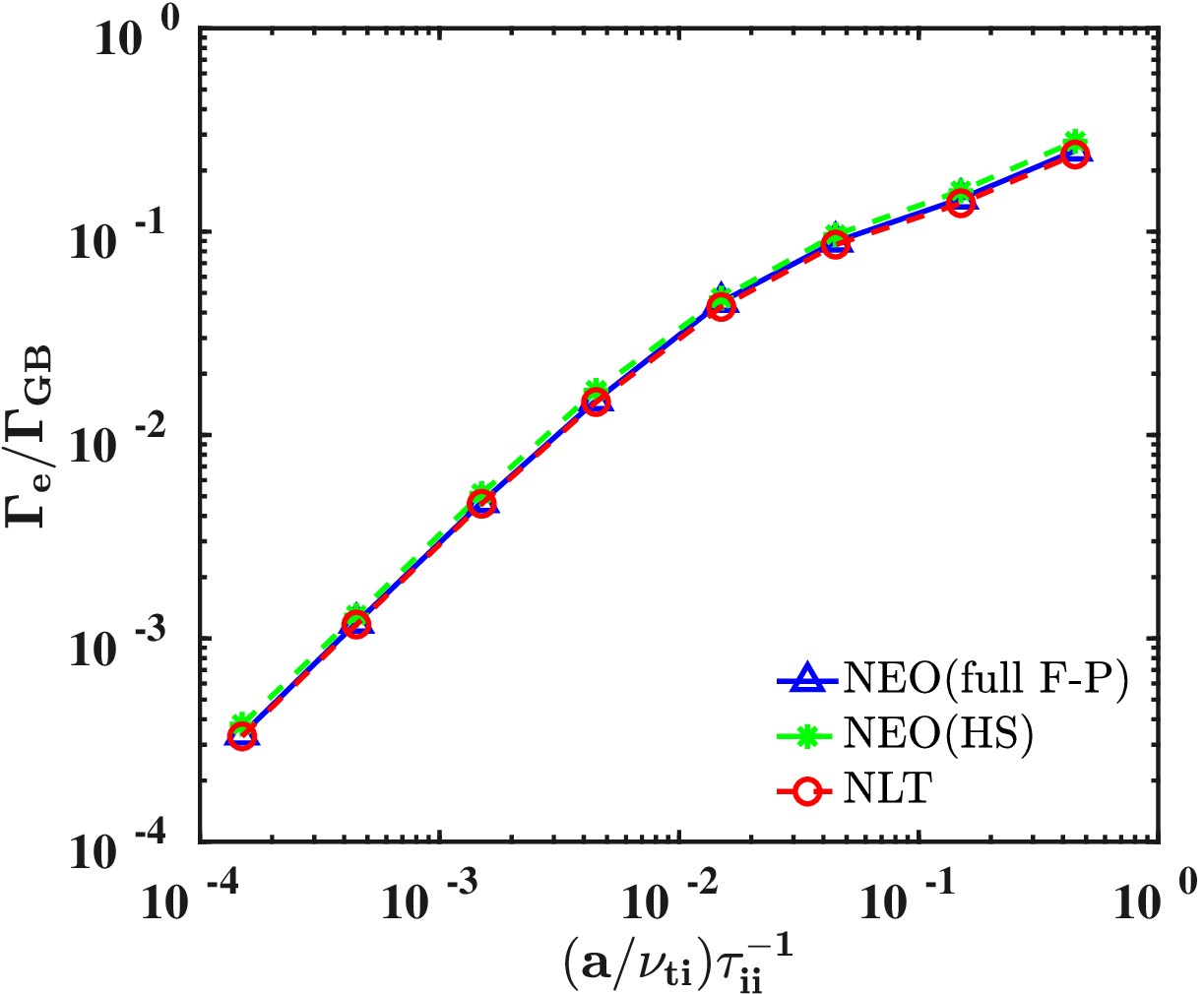}
    \includegraphics[width=0.4\textwidth]{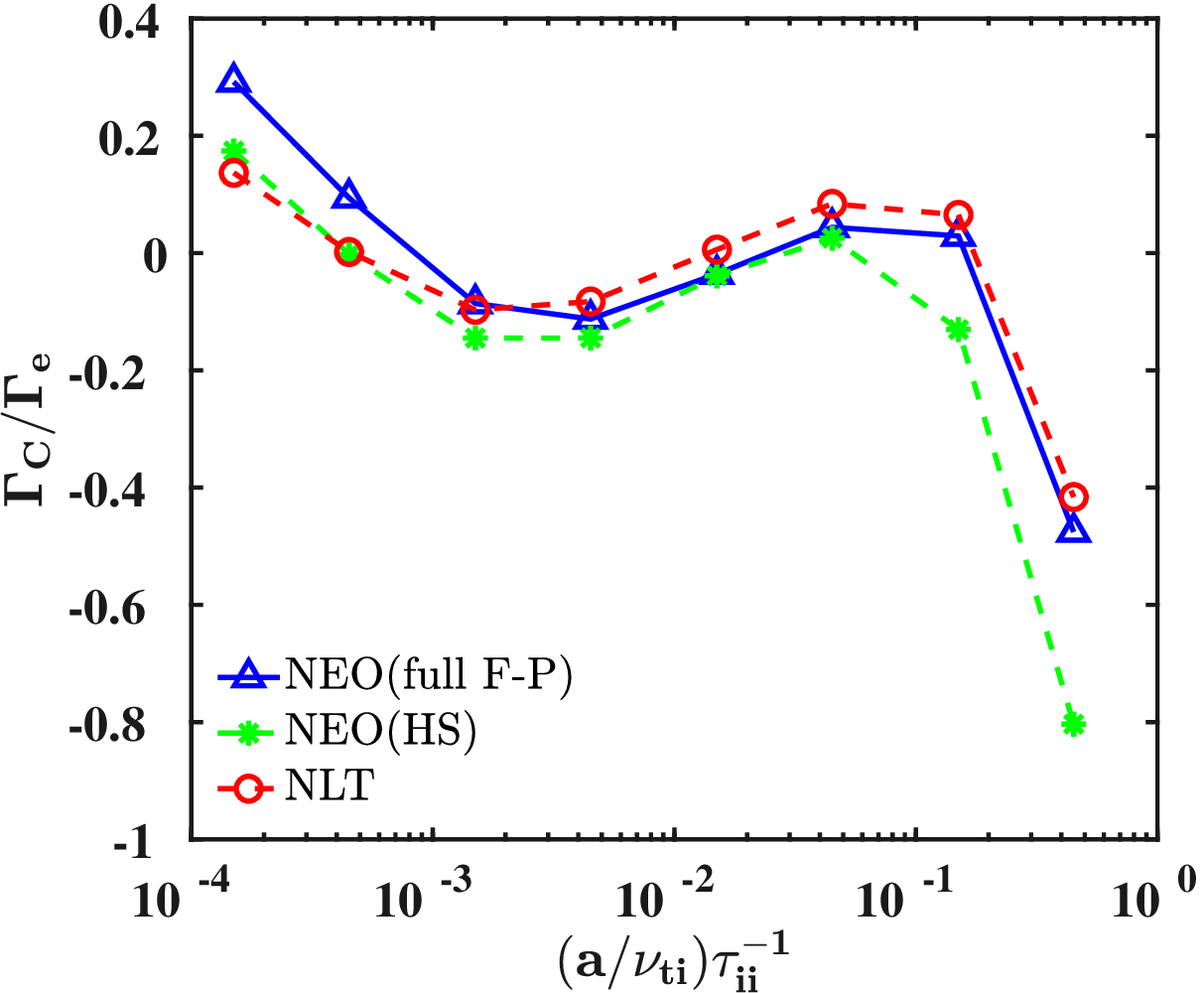}
    \includegraphics[width=0.4\textwidth]{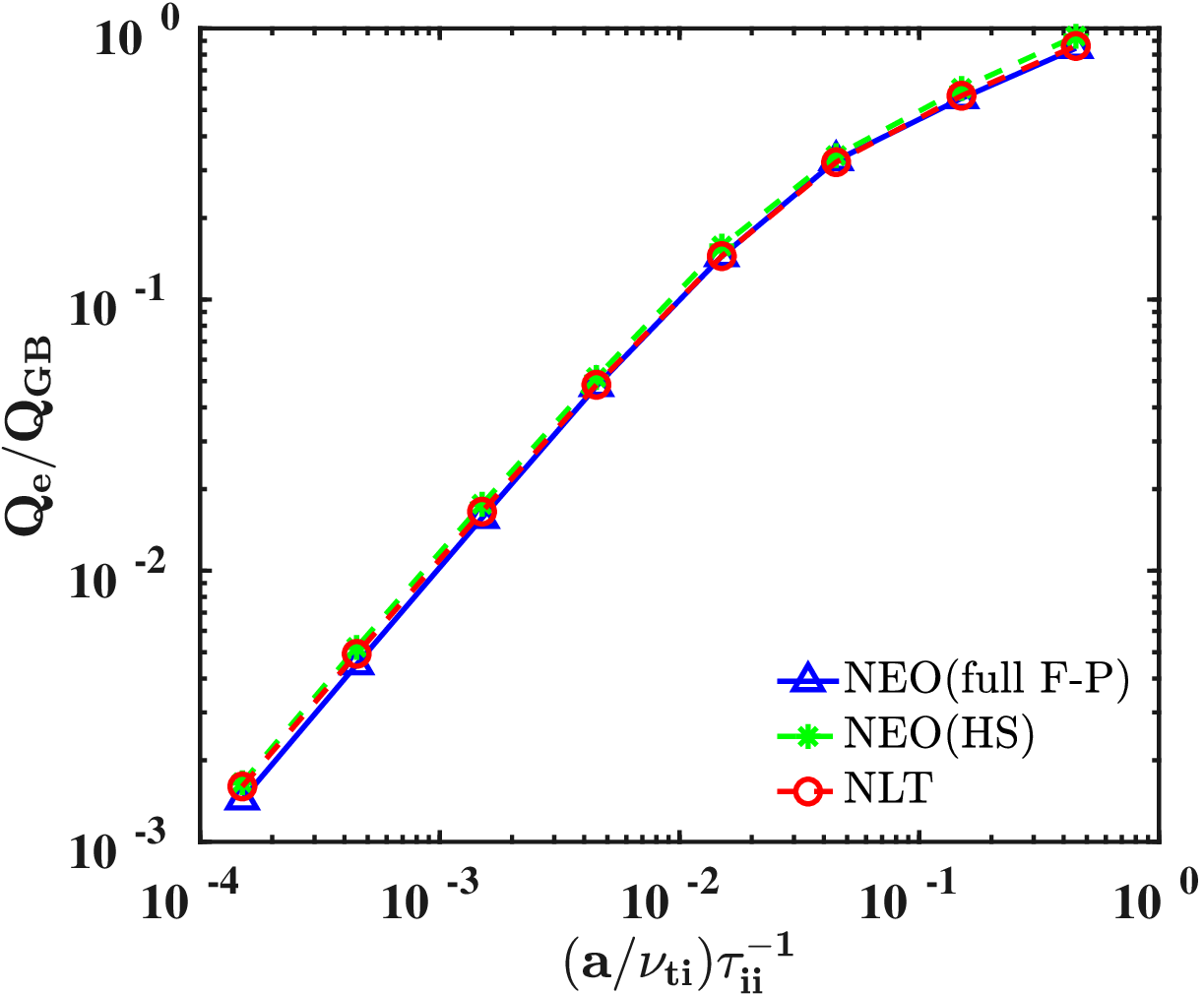}
    \includegraphics[width=0.4\textwidth]{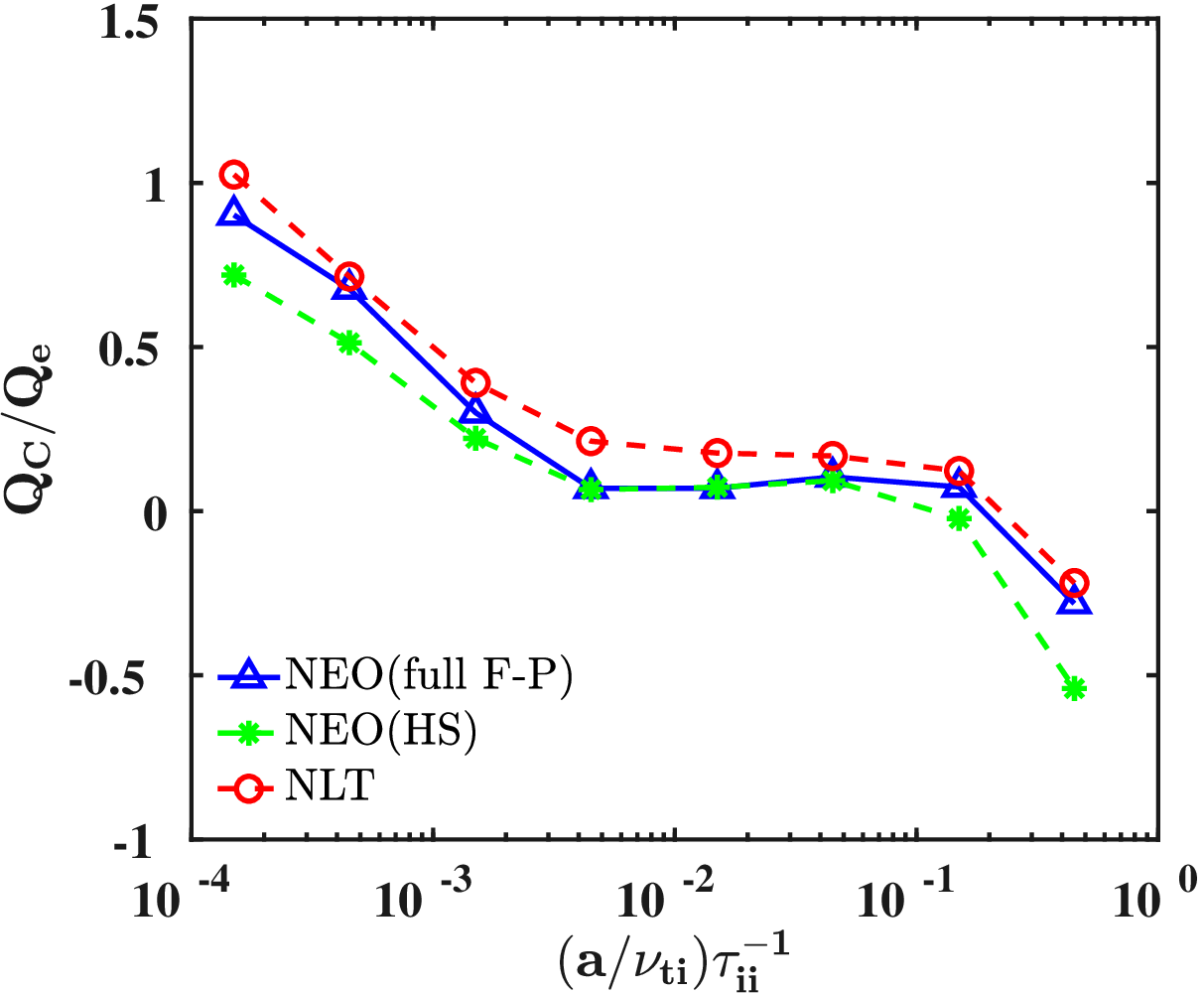}
    \caption{Collisionality dependence of the particle and heat fluxes for electrons and carbon impurities. Blue solid lines denote NEO with the full linearized Fokker-Planck operator, green dashed lines denote NEO with the Hirshman-Sigmar operator, and red dashed lines with symbols denote NLT with the Sugama collision operator.}
    \label{fig:C_G}
\end{figure}

\begin{figure}[t]
\centering
    \includegraphics[width=0.4\textwidth]{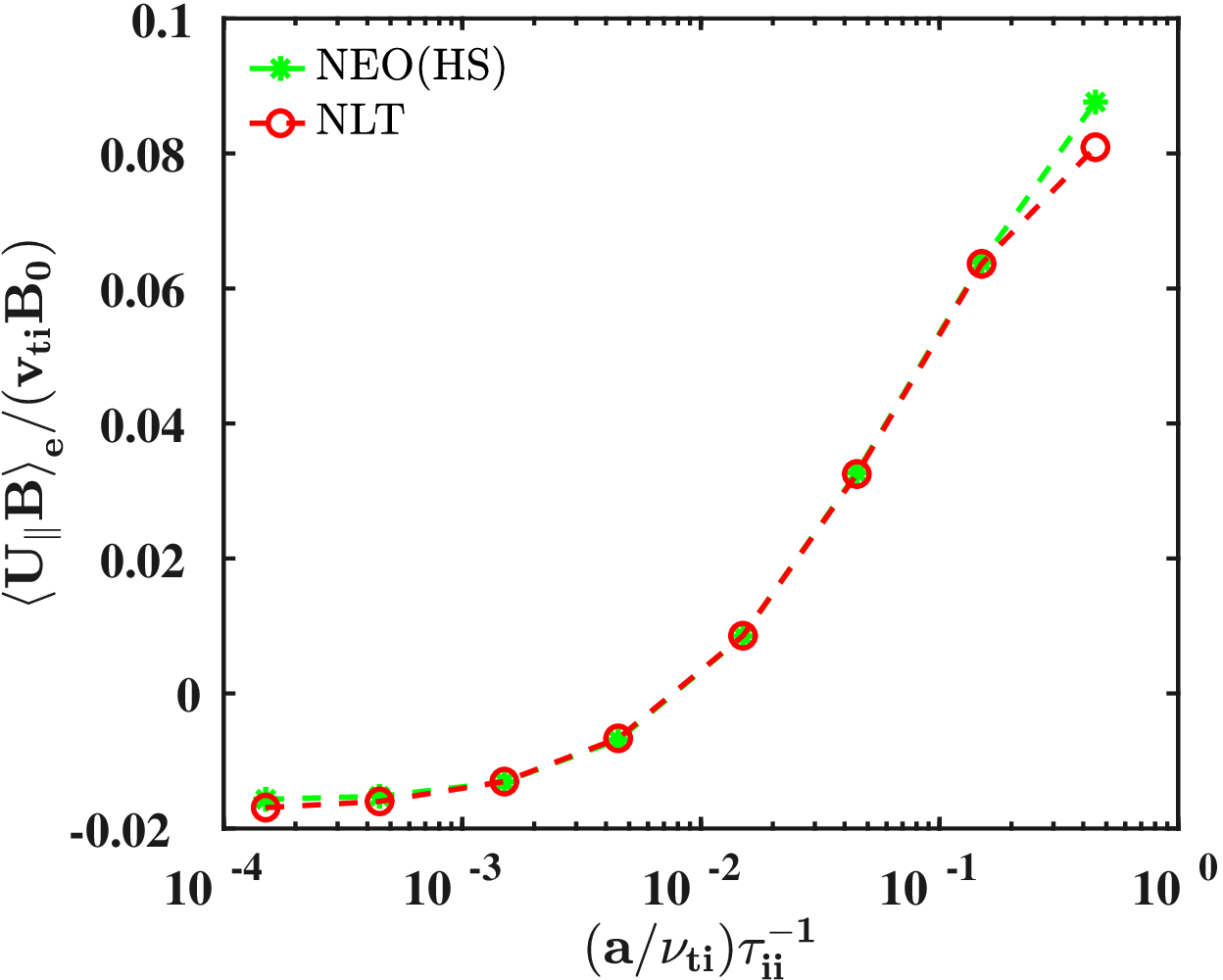}
    \includegraphics[width=0.4\textwidth]{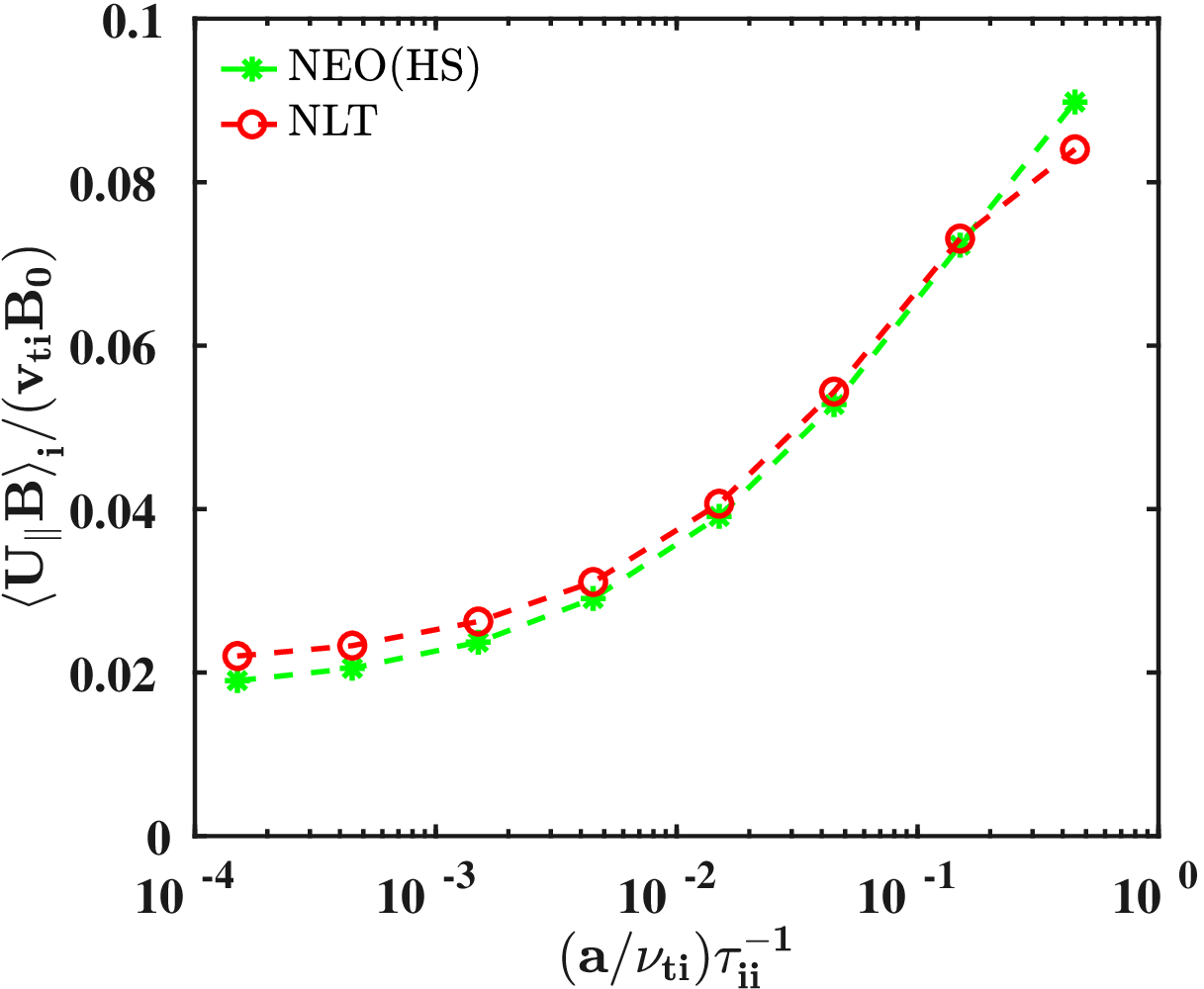}
    \includegraphics[width=0.4\textwidth]{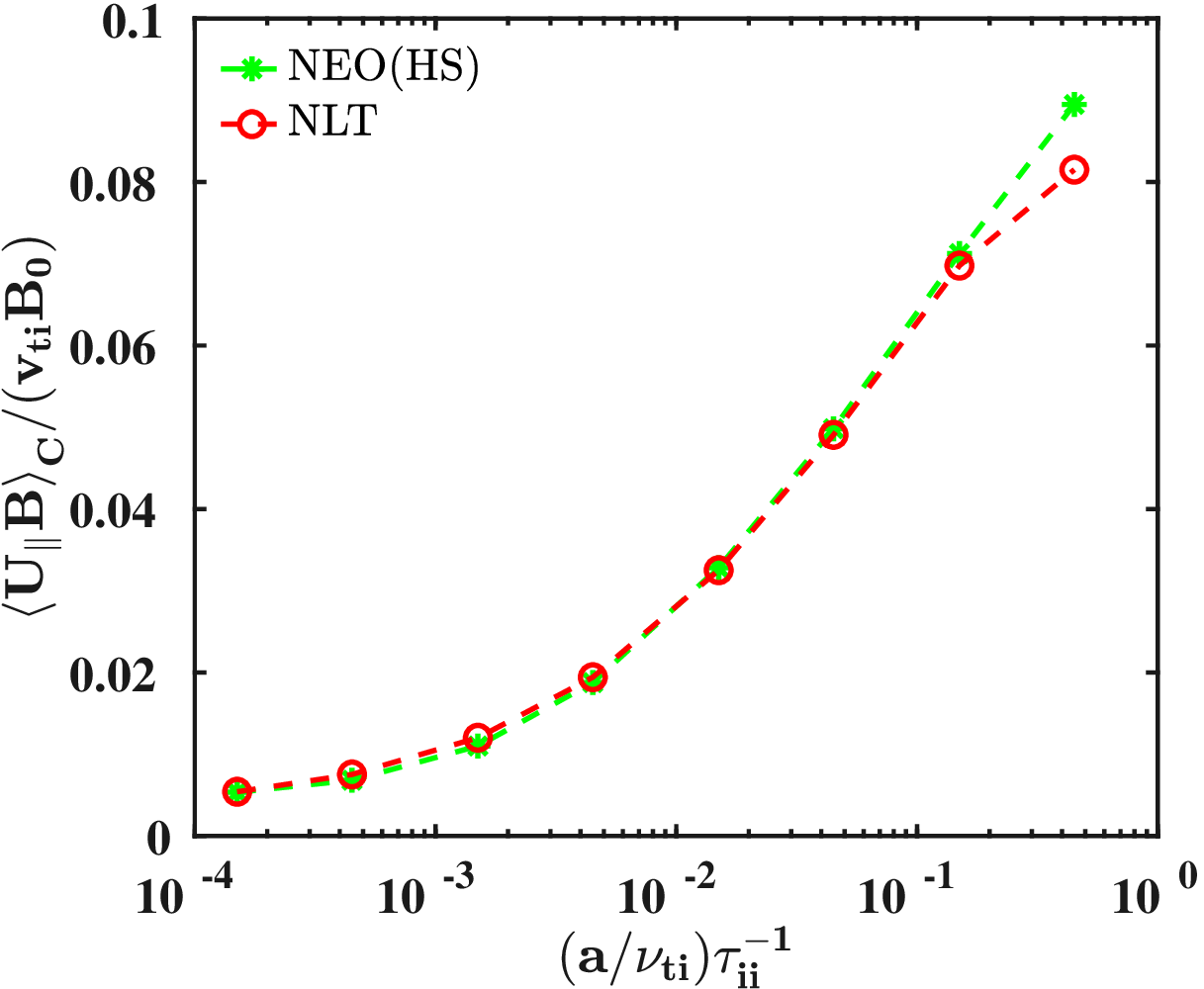}
    \includegraphics[width=0.4\textwidth]{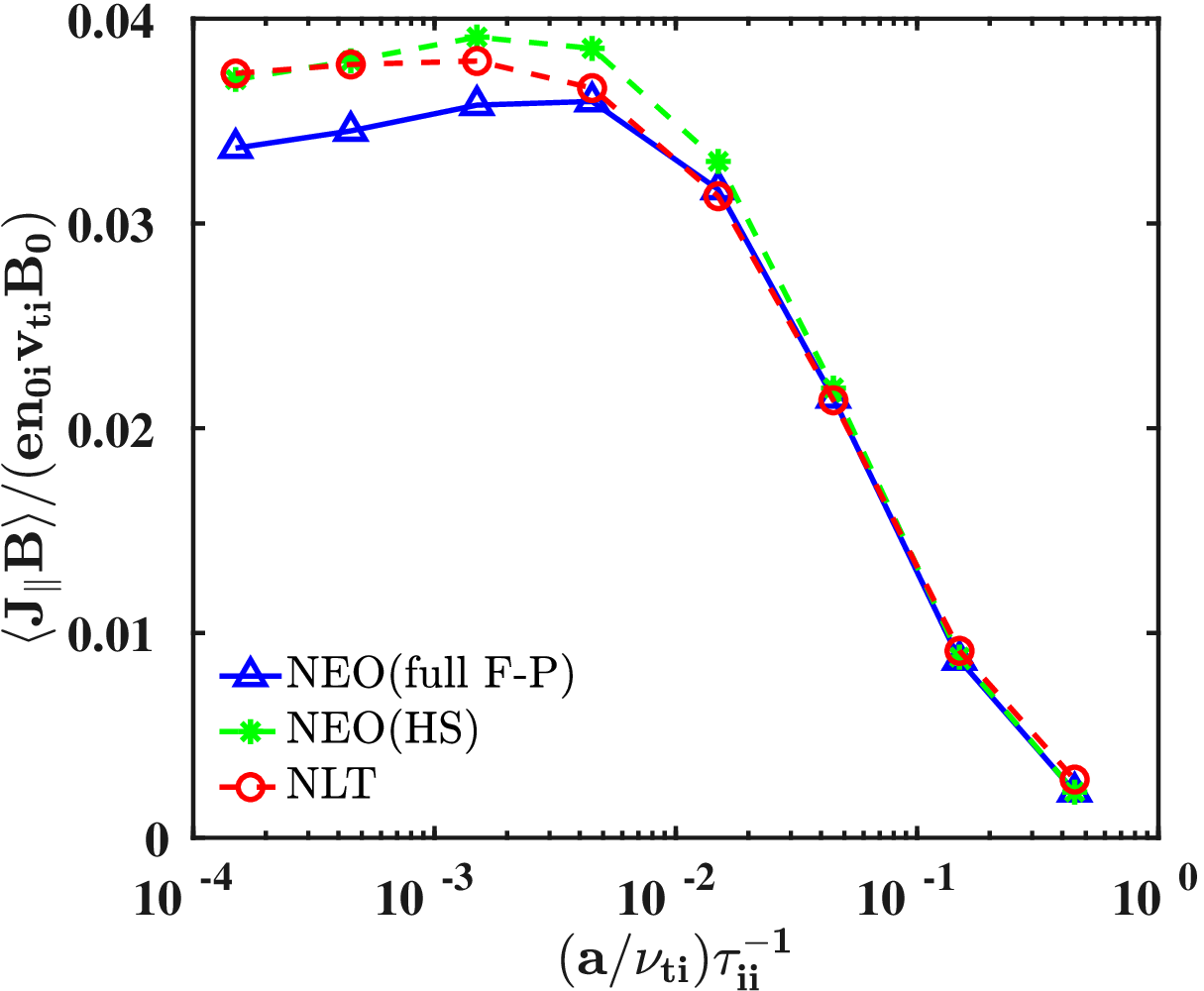}
    \caption{Collisionality dependence of the flux-surface-averaged parallel flows for electrons, deuterium ions, and carbon impurities, together with the total bootstrap current. The line styles are the same as in Fig.~\ref{fig:C_G}.}
    \label{fig:C_U}
\end{figure}

The NLT results obtained with the multi-species Sugama collision operator are compared with NEO calculations using both the full linearized Fokker-Planck operator and the Hirshman-Sigmar model operator. The normalized ion collision frequency is scanned from the banana regime to the Pfirsch-Schluter regime. Figs.~\ref{fig:C_G}-\ref{fig:C_U} summarize the benchmark results for particle fluxes, heat fluxes, flux-surface-averaged parallel flows, and bootstrap current. The particle and heat fluxes from NLT follow the NEO full-Fokker-Planck results over the collisionality range considered. The parallel flows of electrons, deuterium ions, and carbon impurities, together with the total bootstrap current, are also reproduced with good accuracy.

These quantities test different aspects of the collision model and the local neoclassical formulation. The particle fluxes provide a check of multi-species ambipolar transport, the heat fluxes test the energy-dependent response, and the parallel flows and bootstrap current are sensitive to the odd-in-\(v_\parallel\) component of the distribution function and to the momentum-restoring terms in the collision operator. The agreement with NEO therefore validates the implementation of the multi-species Sugama collision model in the local NLT framework.

\section{EAST-relevant tungsten transport and collisional stabilization of TEM}

The multi-species Sugama collision operator implemented in NLT has been validated through the local neoclassical benchmarks in Sec.~3. We now apply the validated collisional NLT framework to EAST-relevant conditions. Recent EAST experiments indicate that discharges with increased tungsten content and larger effective charge can exhibit a reduction of core trapped-electron-mode (TEM)-like density fluctuations. In the present work, we focus on the role of the increased effective collisionality associated with tungsten accumulation and the corresponding increase of \(Z_{\rm eff}\). The neoclassical tungsten transport calculations and the linear TEM simulations are performed independently; their self-consistent coupling is left for future work.

\subsection{Experimental context and simulation setup}

\begin{table}[htbp]
\centering
\caption{Main plasma parameters and effective impurity conditions for EAST Shot~\#$107109$ before and during ICRF heating.}
\label{tab:east_parameters}
\begin{tabular}{lcc}
\toprule
\textbf{Parameter} & \textbf{Before ICRF}& \textbf{During ICRF}\\
\midrule
$R_0$ (m)& \multicolumn{2}{c}{1.85} \\
 $a$ (m)& \multicolumn{2}{c}{0.45} \\
 $I_p$ (MA)& \multicolumn{2}{c}{0.35} \\
 $B$ (T)& \multicolumn{2}{c}{2.5} \\
$n_{e0}$ ($ \text{ m}^{-3}$)& \multicolumn{2}{c}{$\approx 2.4\times 10^{19} $} \\
 \(Z_{\rm eff}\)& \(\approx 2.4\)& \(\approx 3.8\)\\
\(n_W^{\rm eff}\) (\({\rm m}^{-3}\))& \(\approx 3.8\times10^{16}\)& \(\approx 7.7\times10^{16}\)\\
\bottomrule
\end{tabular}
\end{table}
The main plasma parameters and impurity conditions used in the simulations are summarized in Table~\ref{tab:east_parameters}. The experimental input parameters for the simulations are adopted from the analysis reported in Ref.~\cite{zhang2026}. The present study focuses on EAST Shot~\#$107109$, in which the impurity level changes during ICRF heating. After ICRF injection, the effective charge \(Z_{\rm eff}\) increases, accompanied by stronger W-related diagnostic signals and enhanced total radiated power, indicating that the increased impurity content is mainly associated with tungsten~\cite{zhang2026}. For simplicity, only tungsten impurities are retained in the present simulations. The effective tungsten density \(n_W^{\mathrm{eff}}\) listed in Table~\ref{tab:east_parameters} is inferred from the prescribed \(Z_{\rm eff}\) under quasineutrality, rather than from direct experimental measurements of the tungsten density. Despite the increase in impurity radiation after ICRF injection, the ion and electron temperatures are not degraded, and the measured temperature and density profiles show comparable or slightly improved core plasma parameters. Meanwhile, collective Thomson scattering measurements show that the density fluctuation level is significantly reduced at \(k_\theta=10~\mathrm{cm}^{-1}\) and \(20~\mathrm{cm}^{-1}\), both in the inner core and in the outer gradient region. The fluctuation spectrum in the inner core also shifts downward in the frequency range relevant to the observed turbulence. Magnetic-probe measurements and gyrokinetic analyses further indicate that the suppressed fluctuations have TEM-like characteristics and propagate in the electron diamagnetic direction.
  
  The present work focuses on the representative radial position \(r/a=0.6\), where the validated collisional NLT framework is used to examine how the increased effective collisionality associated with the higher \(Z_{\rm eff}\) modifies the linear TEM response. First, impurity-free and tungsten-containing cases with different \(Z_{\rm eff}\) are compared to examine the stabilization trend associated with increased impurity content. Second, for the W-containing case, the collision frequency is artificially scaled while the equilibrium profiles and effective impurity content are kept fixed, so that the role of Coulomb collisions can be isolated. In addition, the neoclassical tungsten particle flux is calculated using the same local profiles to assess the tendency of W transport under these conditions.

\subsection{Collisional stabilization of TEM with increased tungsten concentration}

\begin{figure}[htbp]
\centering
    \includegraphics[width=0.4\textwidth]{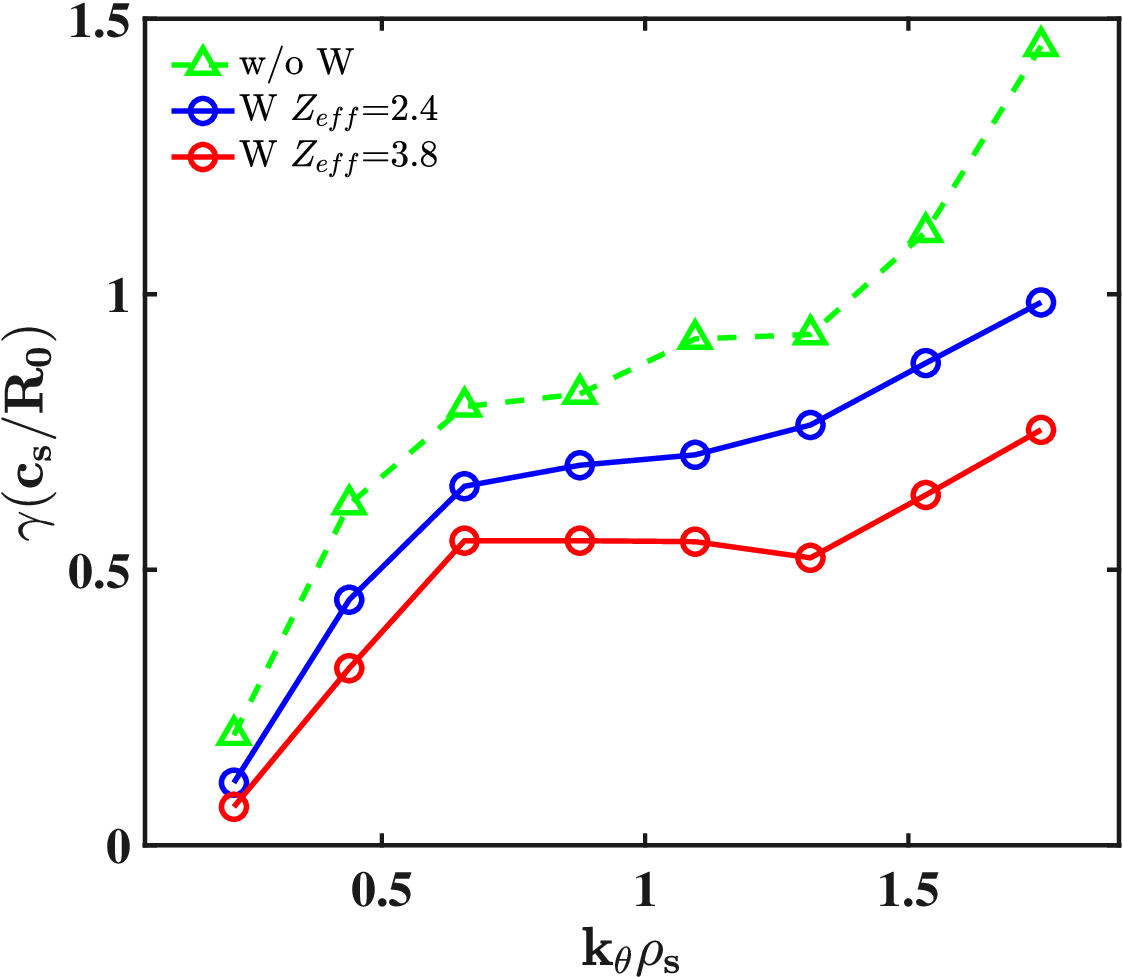}
    \includegraphics[width=0.4\textwidth]{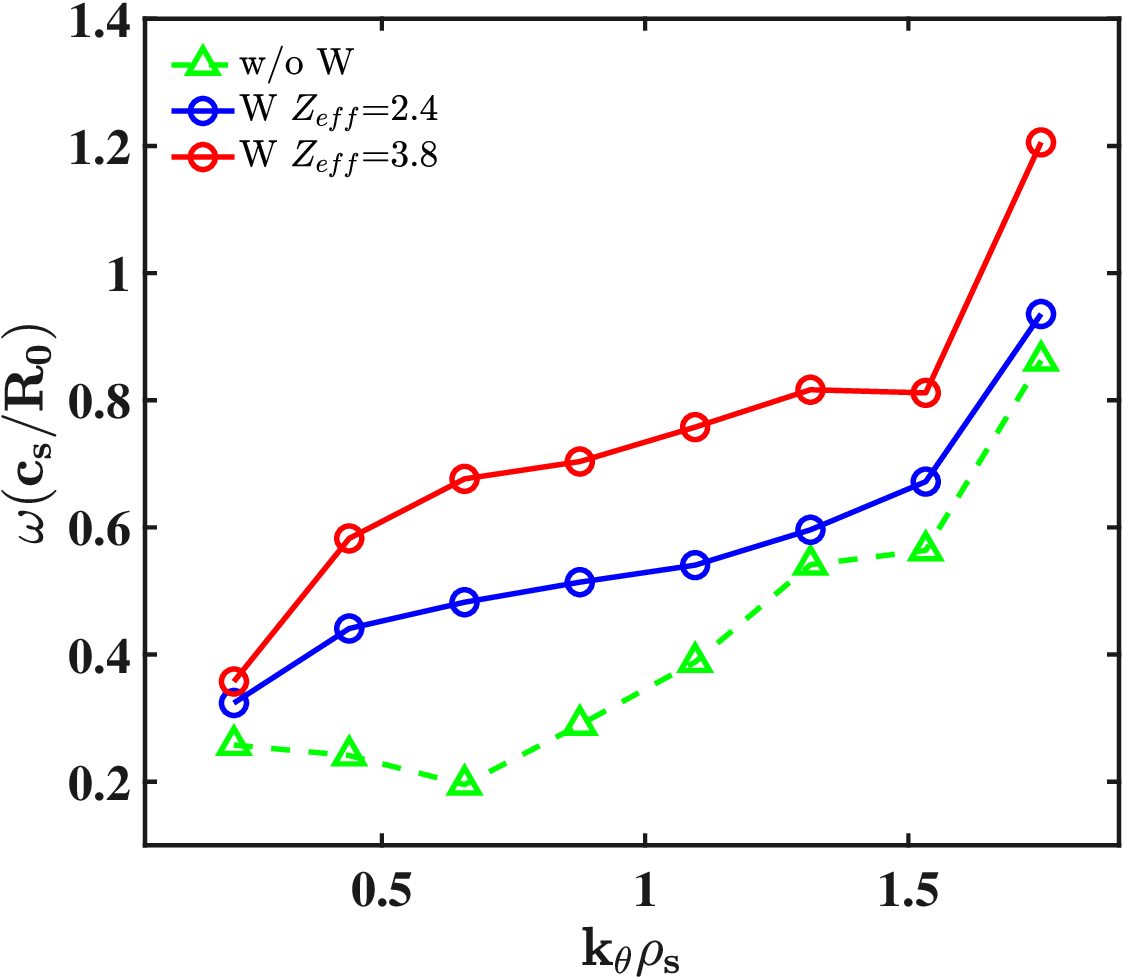}
    \caption{Linear TEM spectrum at \(r/a=0.6\) for impurity-free and W-containing plasmas. The left panel shows the linear growth rate \(\gamma\), and the right panel shows the real frequency \(\omega\). The higher-\(Z_{\rm eff}\) tungsten-containing case gives the strongest reduction of the TEM growth rate.}
    \label{fig:TEM_main}
\end{figure}

We first examine the dependence of the linear TEM spectrum on impurity content and \(Z_{\rm eff}\). Figure~\ref{fig:TEM_main} shows the linear growth rate \(\gamma\) and real frequency \(\omega\), both normalized by \(c_s/R_0\), as functions of \(k_\theta\rho_s\). The impurity-free electron-deuterium plasma is compared with W-containing plasmas at \(Z_{\rm eff}=2.4\) and \(Z_{\rm eff}=3.8\). 

The impurity-free case gives the largest TEM growth rate over the wavenumber range, indicating the strongest trapped-electron drive. When impurity species are included, the growth rate is reduced. A further decrease is observed when \(Z_{\rm eff}\) increases from $2.4$ to $3.8$. Increasing the tungsten impurity concentration, or equivalently the corresponding \(Z_{\rm eff}\), from the lower-impurity case to the higher-impurity case is accompanied by an increase in the effective collisionality. This trend suggests that collisions may play an important role in the experimentally observed TEM suppression. The real frequency is also shifted, indicating a modification of the mode response, while the growth-rate reduction provides the primary evidence for TEM stabilization.

\begin{figure}[t]
\centering
    \includegraphics[width=0.4\textwidth]{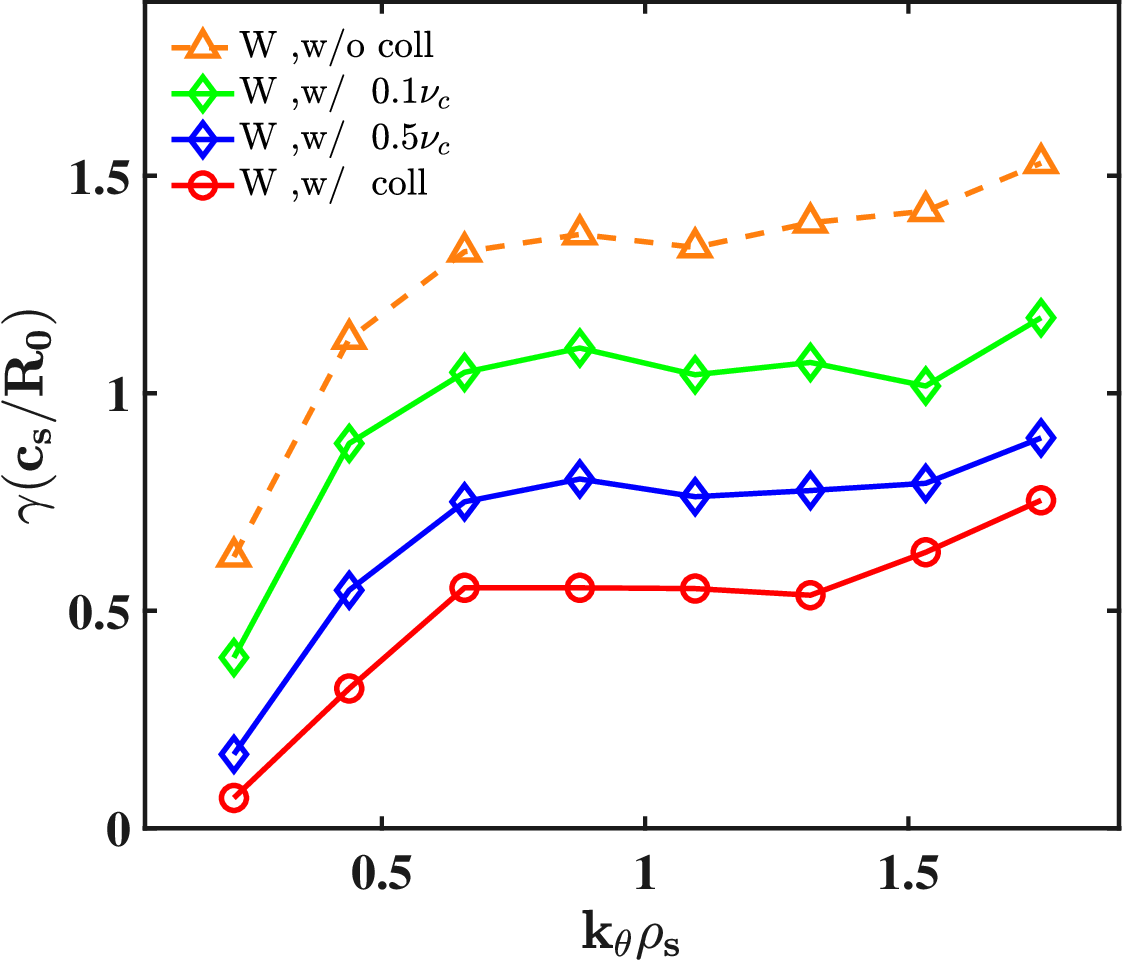}
    \caption{Collisionality scaling of the TEM growth rate for the \(W, Z_{\rm eff}=3.8\) case at \(r/a=0.6\). The collision frequency is scaled from the collisionless limit to \(0.1\nu_c\), \(0.5\nu_c\), and the experimental value \(\nu_c\). The decreasing growth rate with increasing collision frequency demonstrates the stabilizing role of Coulomb collisions.}
    \label{fig:TEM_coll_scan}
\end{figure}

The increase of tungsten concentration can modify the TEM response through several effects, such as main-ion dilution and the increase of the effective collision frequency associated with larger \(Z_{\rm eff}\). To further isolate the collisional contribution, we perform a collision-frequency scaling study for the \(W, Z_{\rm eff}=3.8\) case, as shown in Fig.~\ref{fig:TEM_coll_scan}. In this scan, the equilibrium profiles and tungsten content are kept fixed, while the collision frequency is scaled from the collisionless limit to \(0.1\nu_c\), \(0.5\nu_c\), and the experimental collisionality \(\nu_c\). The growth rate decreases systematically as the collision frequency increases, demonstrating that Coulomb collisions provide an important stabilizing contribution to the TEM under the present EAST-relevant conditions.

Physically, the stabilization can be understood as a collisional weakening of the trapped-electron response. Pitch-angle scattering transfers part of the trapped-electron population into passing orbits and broadens the trapped-passing boundary region, thereby reducing the resonant trapped-electron precessional drive of the TEM \cite{TEM2015,XIAO2025}. In this sense, the increase of tungsten concentration can affect TEM stability indirectly through the associated increase of \(Z_{\rm eff}\) and effective electron collisionality.

\begin{figure}[t]
\centering
    \includegraphics[width=0.4\textwidth]{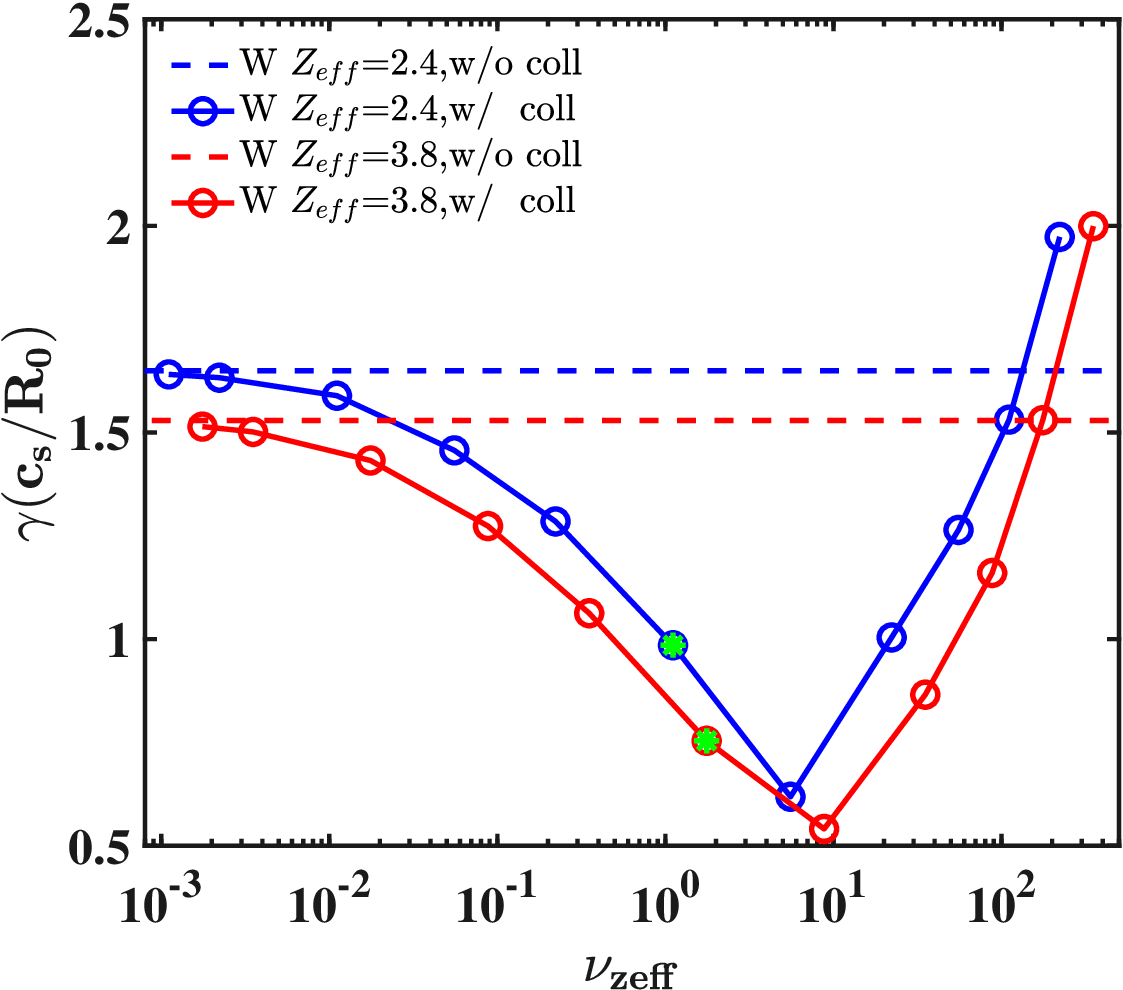}
    \includegraphics[width=0.4\textwidth]{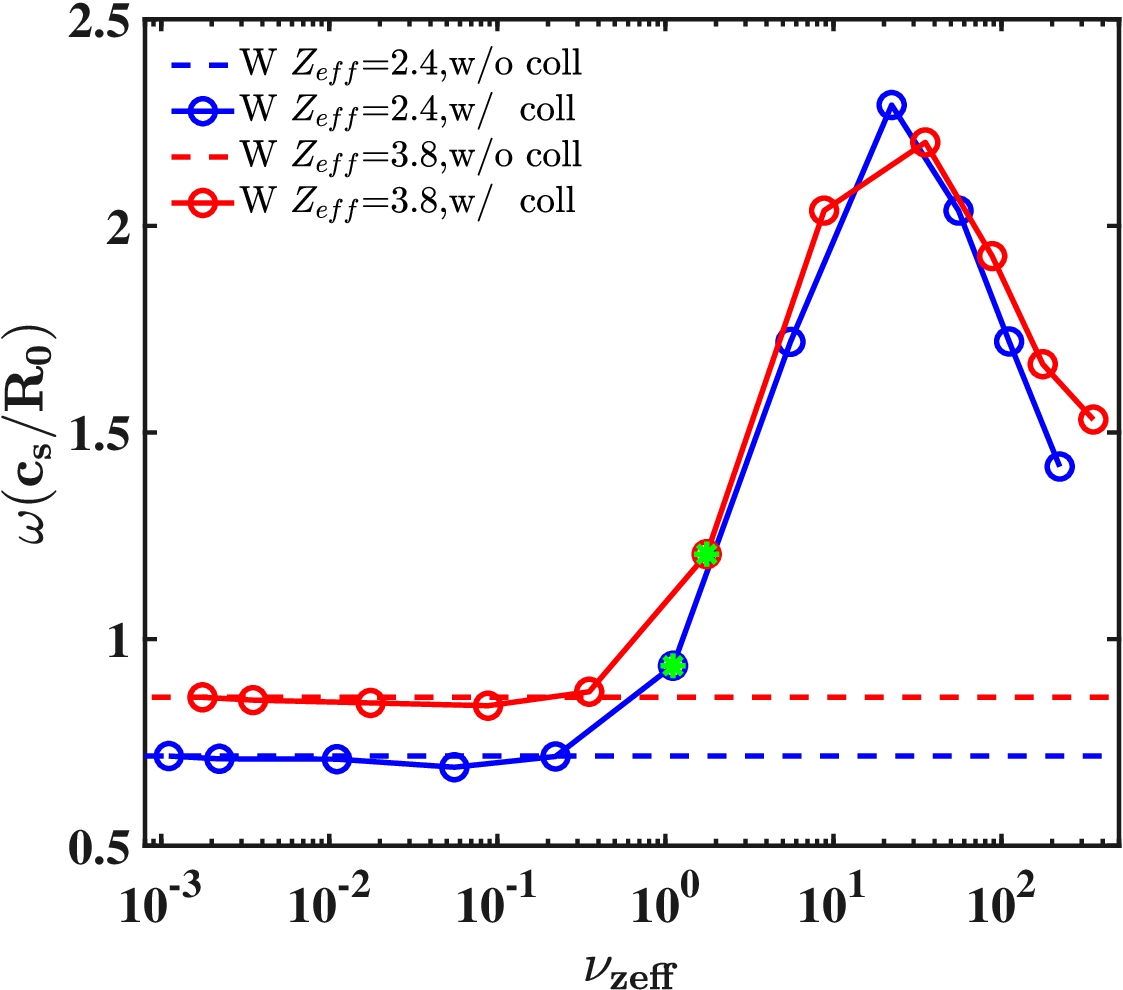}
    \caption{Dependence of the linear growth rate and real frequency on the effective collision frequency for tungsten-containing plasmas with \(Z_{\rm eff}=2.4\) and \(Z_{\rm eff}=3.8\). The dashed horizontal lines denote the collisionless reference values, and the green markers indicate the experimental values. The growth rate first decreases with increasing collisionality and then increases at high collisionality, accompanied by a significant frequency shift.}
    \label{fig:TEM_coll_n80}
\end{figure}
At the wavenumber corresponding approximately to \(k_\theta=10~{\rm cm}^{-1}\), Fig.~\ref{fig:TEM_coll_n80} further shows the dependence of the linear growth rate and real frequency on the effective collision frequency for tungsten-containing plasmas with \(Z_{\rm eff}=2.4\) and \(Z_{\rm eff}=3.8\). The dashed horizontal lines denote the corresponding collisionless reference values, and the green markers indicate the experimental values. In the low- to intermediate-collisionality range, the growth rate decreases with increasing collision frequency for both \(Z_{\rm eff}\) cases, confirming the collisional stabilization trend. The \(Z_{\rm eff}=3.8\) case remains more stable over most of the experimentally relevant range, consistent with the stronger TEM suppression observed in the higher-W discharge. At higher collisionality, the growth rate increases again after reaching a minimum, accompanied by a clear shift of the real frequency. This non-monotonic behavior suggests that the mode character may change when the collisionality becomes sufficiently large. Therefore, the experimentally relevant stabilization should be associated with the decreasing-growth-rate regime of the scan, while the high-collisionality upturn should be interpreted cautiously as a possible transition toward a dissipative TEM-like response rather than as a continuation of the same collisionless TEM branch.

\subsection{Neoclassical tungsten particle transport}

\begin{figure}[t]
\centering
    \includegraphics[width=0.4\textwidth]{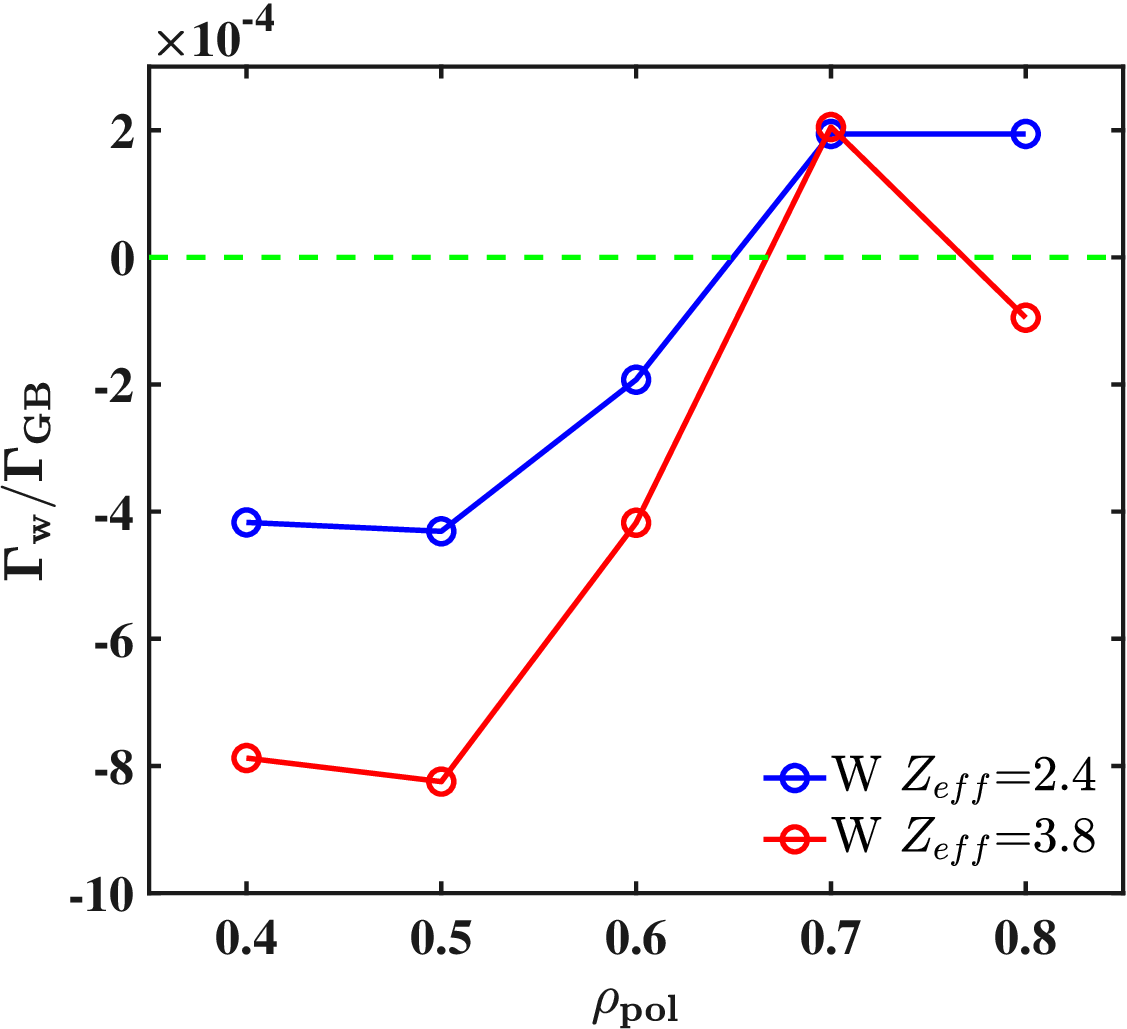}
    \caption{Radial profile of the neoclassical tungsten particle flux normalized by the gyro-Bohm particle flux for \(W, Z_{\rm eff}=2.4\) and \(W, Z_{\rm eff}=3.8\). The dashed horizontal line denotes the zero-flux level. Negative values correspond to inward tungsten transport, while positive values indicate outward transport.}
    \label{fig:flux_n80}
\end{figure}

In addition to the linear TEM stability analysis, we calculate the neoclassical tungsten particle flux under the same EAST-relevant conditions. Figure~\ref{fig:flux_n80} shows the radial profile of the tungsten particle flux normalized by the gyro-Bohm particle flux for \(W, Z_{\rm eff}=2.4\) and \(W, Z_{\rm eff}=3.8\). Negative values correspond to inward tungsten transport, while positive values indicate outward transport.

In the inner radial region, the tungsten particle flux is predominantly inward, and the inward flux is stronger for the \(Z_{\rm eff}=3.8\) case. This indicates that higher tungsten content and larger effective charge can enhance the local neoclassical tendency toward tungsten accumulation. These results show that neoclassical tungsten transport is sensitive to local profile gradients and can vary substantially across the minor radius.

\begin{figure}[t]
\centering
    \includegraphics[width=0.4\textwidth]{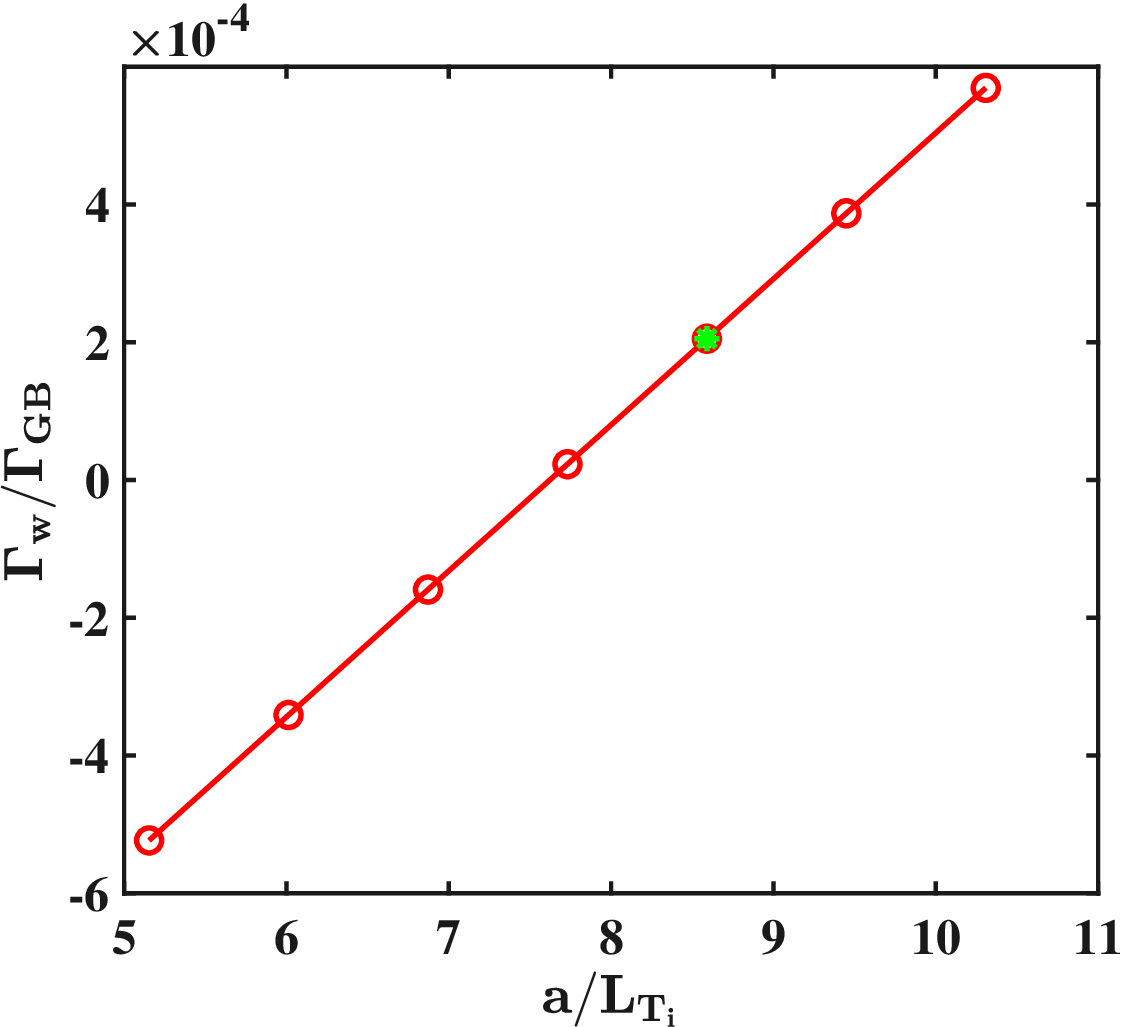}
    \caption{Dependence of the neoclassical tungsten particle flux on the main-ion temperature gradient \(a/L_{T_i}\) at \(Z_{\rm eff}=3.8\) and \(\rho_{\rm pol}=0.7\). The green marker denotes the experimental value. Negative values correspond to inward tungsten transport, while positive values indicate outward transport. The sign reversal of \(\Gamma_W\) reflects the competition between the inward neoclassical pinch and the outward ion-temperature-gradient-driven temperature-screening effect. }
    \label{fig:flux_Ti}
\end{figure}
Around \(\rho_{\rm pol}\approx0.7\), the tungsten particle flux changes sign and becomes outward. To clarify the physical origin of this outward flux, we perform an additional scan of the main-ion temperature gradient \(a/L_{T_i}\) at \(Z_{\rm eff}=3.8\) and \(\rho_{\rm pol}=0.7\), as shown in Fig.~\ref{fig:flux_Ti}. When \(a/L_{T_i}\) is small, \(\Gamma_W/\Gamma_{\rm GB}<0\), indicating inward tungsten transport. As \(a/L_{T_i}\) increases, the tungsten flux increases and eventually becomes positive. The experimental point, marked by the green symbol, lies in the outward-flux regime. This trend suggests that the outward tungsten flux near \(\rho_{\rm pol}=0.7\) is driven by the main-ion-temperature-gradient-induced temperature-screening effect, which can overcome the inward neoclassical pinch when the ion temperature gradient is sufficiently large.

Since the present calculation uses prescribed local profiles and does not include self-consistent impurity-density evolution, the computed tungsten flux should be interpreted as a local neoclassical tendency rather than a predictive impurity accumulation simulation. Taken together, the linear TEM simulations and the neoclassical tungsten transport calculations provide a mechanism-oriented interpretation of the EAST-relevant cases considered here. The TEM calculations show that the increase of effective collisionality associated with higher tungsten content can reduce the linear TEM growth rate in the experimentally relevant parameter range, while the neoclassical calculations indicate that the tungsten flux is inward over the inner part of the simulated radial domain and can become outward near \(\rho_{\rm pol}\approx0.7\) due to ion-temperature-gradient-driven temperature screening. Nonlinear TEM saturation, self-consistent tungsten profile evolution, and the coupling between turbulent and neoclassical transport are left for future studies.

\section{Summary and discussion}

A local neoclassical transport module has been developed and validated in the semi-Lagrangian gyrokinetic code NLT for multi-species collisional plasmas. The module incorporates the linearized multi-species Sugama collision operator and supports two complementary approaches: an initial-value relaxation method with composite substep source integration, and a direct steady-state solver for the stationary neoclassical response. The former is consistent with the time-evolution framework of NLT turbulence simulations, while the latter avoids long-time relaxation and repeated characteristic tracing.

Benchmark tests against the Eulerian neoclassical code NEO show that the local NLT module reproduces particle fluxes, heat fluxes, \(B\)-weighted parallel flows, and bootstrap current over a broad collisionality range. The electron-ion benchmarks verify both the initial-value and direct steady-state solvers, while the three-species carbon-impurity benchmark further validates the multi-species collision coupling, including inter-species momentum and energy exchange. The convergence and ambipolarity tests demonstrate that the composite substep source-integration scheme improves the accuracy of orbit-integrated neoclassical source terms, especially for fast electron motion and low-collisionality cases.

As EAST-relevant applications, the validated framework has been used to study tungsten neoclassical transport and collisional stabilization of core TEM. The results show that increased tungsten content, through the associated increase of \(Z_{\rm eff}\) and effective collisionality, can reduce the linear TEM growth rate in the experimentally relevant range, while the neoclassical tungsten flux is inward over the inner part of the simulated radial domain and can become outward near \(\rho_{\rm pol}\approx0.7\) due to ion-temperature-gradient-driven temperature screening. These results should be interpreted as a mechanism-oriented application of the validated collisional NLT framework, since nonlinear TEM saturation, turbulent impurity transport, radiation feedback, and self-consistent tungsten profile evolution are not included. Future work will extend the framework to nonlinear multi-species simulations and coupled studies of neoclassical and turbulent impurity transport.

   \section*{Acknowledgment}
   This work was supported by the Strategic Priority Research Program of the Chinese Academy of Sciences under Grant No. XDB0790201,  the National Natural Science Foundation of China under Grant No. 12405275, the National MCF Energy R\&D Program of China under Grant Nos. 2019YFE03060000, 2022YFE03060004 and 2022YFE03190200. Part of the numerical calculations in this work were performed on ShenMa High Performance Computing Cluster in Institute of Plasma Physics,  Chinese Academy of Sciences,  Hefei advanced computing center and National Supercomputer Center in Tianjin.

\bibliography{PST_references}

\end{document}